\documentclass[aps,prd,showpacs,10pt]{revtex4-1}

\usepackage{amsfonts}

\begin{document}

\title{Conditions for the  emergence of gauge bosons from spontaneous Lorentz
symmetry breaking \\
}
\author{C. A. Escobar $^{1}$ and L. F. Urrutia $^{1}$ }
\affiliation{$^{1}$ Instituto de Ciencias Nucleares, Universidad Nacional Aut{\'o}noma de
M{\'e}xico, A. Postal 70-543, 04510 M{\'e}xico D.F., M{\'e}xico }

\begin{abstract}
The emergence of gauge particles (e.g., photons and gravitons)
as Goldstone bosons arising from spontaneous symmetry breaking is
an interesting hypothesis which would provide a dynamical setting for the
gauge principle. We investigate this proposal in the framework of a general 
SO$(N)$ non-Abelian Nambu model (NANM), effectively providing spontaneous
Lorentz symmetry breaking in terms of the corresponding Goldstone bosons.
Using a nonperturbative Hamiltonian analysis, we prove that the SO$(N)$ 
Yang-Mills theory is equivalent to the corresponding NANM, after both current
conservation and  the Gauss laws are imposed as initial conditions
for the latter. This equivalence is independent of any gauge fixing in the
YM theory. A substantial conceptual and practical improvement in the
analysis arises by choosing a particular parametrization that solves the
nonlinear constraint defining the NANM. This choice allows us to show that
the relation between the NANM canonical variables and the corresponding ones
of the YM theory, $A_{i}^{a}$ and $E^{bj}$, is given by a canonical
transformation. In terms of the latter variables, the NANM Hamiltonian has
the same form as the YM Hamiltonian, except that the Gauss laws do not arise
as first-class constraints. The dynamics of the NANM further guarantees that
it is sufficient to impose them only as initial conditions, in order to
recover the full equivalence. It is interesting to observe that this
particular parametrization exhibits the NANM as a regular theory, thus
providing a substantial simplification in the calculations.
\end{abstract}

\maketitle

\section{Introduction}

\label{intro} The possible interpretation of gauge particles (e.g., photons
and gravitons) as Goldstone bosons (GBs) arising from some
spontaneous symmetry breaking, dates back to the pioneering works of%
\cite{RUSOS} and \cite{Bjorken,Guralnik}. The former used the standard coset
construction of the effective theory \cite{Coset}, in which the spontaneous Lorentz symmetry breaking (SLSB) is
realized nonlinearly in terms of the GBs, with the matter fields transforming
linearly under the unbroken subgroup. Their conclusion was that any gauge
theory is a theory of some spontaneously broken symmetry. A recent
application of the coset construction to general relativity was reported in Ref. 
\cite{Tomboulis}.

On the other hand, Refs. \cite{Bjorken,Guralnik} considered the explicit
case of spontaneous Lorentz symmetry breaking to produce the photon
field as the condensate arising from a self-coupled four-fermion model,
following similar steps as those developed in Ref. \cite{Nambu_Jona-Lasinio}
to describe the superconductor solutions in field theory. Similar ideas have
been revisited for photons and also have been extended to gravitons in Refs. 
\cite{Phillips,Ohanian,Kraus}.

An alternative approach was proposed by Nambu in Ref. \cite{Nambu-Progr},
where the emphasis was shifted to the description of the SLSB system
(QED, in this case) only in terms of the GB
degrees of freedom (d.o.f.), which were introduced via a nonlinear constraint,
similarly to the nonlinear sigma model description of pion
interactions. Such a $\sigma $-QED is defined by the Maxwell's Lagrangian plus
the constraint $A_{\mu }A^{\mu }=n^{2}M^{2}$, which is to be substituted into
the Lagrangian. Here, $n_{\mu }$ is a properly oriented constant Lorentz
vector, while $M$ is the proposed scale associated with the SLSB. This
constraint can be understood as providing a nonzero vacuum expectation value $%
\langle A_{\mu }\rangle=n_{\mu }M$. Nevertheless, the goal in this model was
to show that it is in fact equivalent to standard QED, instead of yielding a
physical violation of the Lorentz symmetry. This equivalence was manifest up
to the tree-level calculations studied in Ref. \cite{Nambu-Progr}. Later on,
these calculations were extended to some processes at the one-loop level,
with identical results: all contributions arising from the SLSB sector of
the model canceled out, yielding the standard QED results \cite%
{Azatov-Chkareuli}. The $\sigma $-QED model has been further studied \cite{VENT1} and
extended to the non-Abelian \cite{VENT2,JLCH1, JLCH2,JLCH3} and gravitational
cases \cite{JLCH4}, which we will generically call generalized
Nambu models. General conditions regarding how the gauge symmetries were
recovered from the corresponding SLSB models are worked out in  \cite%
{JLCH5,JLCH6,JLCH7}.

Perturbative calculations in the non-Abelian case show again that, to the
order considered, all SLSB contributions to physical processes cancel out,
yielding an equivalence with the starting Yang-Mills (YM) theory, in complete analogy
with the Abelian case. This fact has been interpreted by stating that the
corresponding nonlinear constraint, which defines each Nambu model, can be
interpreted  as just a gauge choice in the associated Abelian or non-Abelian
gauge theory. This would lead to an equivalence between the Nambu model and
the corresponding gauge theory in a fixed gauge. Nevertheless, this
statement requires some qualifications. (i) To begin with, the number of d.o.f.
of the Nambu model is larger than that of the corresponding gauge theory,
which can be understood because the former has lost gauge invariance.
(ii) Fixing the gauge in any gauge theory requires the introduction of
ghost particles (via the Becchi-Rouet-Stora-Tyutin  procedure, for example Ref.  \cite{BRST}) which play
a fundamental role as internal particles in calculating physical processes.
Thus, in order to establish the proposed equivalence, one would need to study
the contributions of the ghosts to physical processes. A possible decoupling
of them is by no means evident, especially due to the nonlinear character of
the proposed gauge fixing. The first point (i) has been taken into account
in most previous works and was  emphasized in Ref.  \cite{Urru-Mont}. The general
statement, phrased in different ways in different papers,
is that the Nambu model is equivalent to the corresponding gauge theory only
after current conservation together with the Gauss laws have been imposed as
initial conditions, since the dynamics of the Nambu model preserves their
conservation for all times. The second point (ii) has not been considered at
all and will be dealt with in a separate publication, for the case of the
Abelian Nambu model \cite{Urru-Escobar}.

Recently, the study of possible observable violations of Lorentz invariance
has attracted considerable attention, from both the experimental and
theoretical points of view. Explicit Lorentz symmetry violation is found to
be incompatible with the Bianchi identities \cite{Bianchi_iden} and
therefore this approach is not consistent with general relativity, unlike
SLSB, where this issue does not occur. The construction of the Standard Model
extension performed by Kosleteck\'y and collaborators \cite%
{Colladay-Kostelecky} is a framework in which Lorentz violation is
considered as arising from a spontaneous symmetry breaking in a more
fundamental theory. A distinguished class of models in that framework are
the so-called bumblebee models, which are tensor theories exhibiting
physical SLSB. They include GB modes, and depending on the explicit form of
the theory they have additional modes and constraints. These models have been
thoroughly investigated in relation to electrodynamics \cite{BLUHM_1,
HERNASKI} and gravity \cite{BLUHM_3, BLUHM_2,KOS_POT, CARROLL}. As a matter
of fact, generalized Nambu models can be thought of as very particular cases
of bumblebee models, where the non-Goldstonic d.o.f. of the latter are frozen,
leaving only the GB excitations.

In the present paper we generalize to the non-Abelian case and improve the
nonperturbative Hamiltonian analysis developed for the Abelian Nambu model
(ANM) in Ref. \cite{Urru-Mont}. The non-Abelian Nambu model (NANM)
associated to the group SO$(N)$ is defined by the YM Lagrangian 
\begin{equation}
\mathcal{L}(A_{\mu }^{a})=-\frac{1}{4}F_{\mu \nu }^{a}F^{a\mu \nu }-J^{a\mu
}A_{\mu }^{a},  \label{YML}
\end{equation}%
plus the condition 
\begin{equation}
A_{\mu }^{a}A^{a\mu }=n^{2}M^{2},\qquad M^{2}>0,\quad \mu =0,1,2,3,\quad
a=1,...,N,  \label{NANMC}
\end{equation}%
which is to be solved and substituted into the YM Lagrangian. Our goal is to
determine which additional conditions have to be imposed upon the NANM in
order that its Hamiltonian reduces to that of the YM theory. No discussion
is provided of the possible perturbative equivalence of the so-corrected Nambu model and the YM theory in the fixed nonlinear gauge.

We proceed via the following steps. (i) We start by constructing the
Hamiltonian for each type of NANM (depending on $n^{2}$ it can be time-like,
space-like, or light-like) in terms of the corresponding canonical variables
for each case. (ii) We show that these Hamiltonians are related via a
canonical transformation to a Hamiltonian that has the same form as the YM
Hamiltonian in the standard variables $A_{i}^{a},E_{i}^{a},\,\,i=1,2,3$,
except for the fact that the Gauss laws $\Omega ^{a}=0$ do not appear as
constraints; nevertheless, the canonical transformation leads to the correct
brackets between the canonical variables $A_{i}^{a},E_{i}^{a}$ arising from
the canonical algebra of each NANM. (iii) We prove that the NANM dynamics
preserves the evolution of $\Omega ^{a}$ in such a way that it
guarantees that the imposition of $\Omega ^{a}=0$ for some initial time
leads to $\Omega ^{a}(t)=0$ for all times. In this way, enforcing the Gauss
laws as first-class Hamiltonian constraints at some initial time makes the
NANM equivalent to the corresponding YM theory in a nonperturbative way and
independently of any gauge fixing. Consistency with the NANM dynamics avoids
the generation of additional constraints.

The present paper is organized as follows. In Sec. II we consider the specific
case of the space-like NANM (SL-NANM) by solving the nonlinear constraint (%
\ref{NANMC}) in terms of $A_{\mu =3}^{a=1}$ and starting with the remaining $%
4N-1$ d.o.f. per point in coordinate space. The Lagrangian equations of motion
are obtained and the canonical momenta together with the canonical
Hamiltonian are subsequently constructed. The standard variables of the YM
theory $A_{i}^{a}$ and $E^{bj},i=1,2,3$ are written in terms of the
canonical variables of the SL-NANM and their induced algebra is calculated,
which is summarized in Appendix B. The analysis of the SL-NANM in terms
of the Dirac method reveals that this model has additional second-class
constraints, which are further imposed strongly by introducing Dirac brackets to obtain the d.o.f. of the reduced phase space, together with their
algebra. The induced Dirac-brackets algebra for the variables $A_{i}^{a}$
and $E^{bj}$ are also calculated. The final extended Hamiltonian for the
SL-NANM, rewritten in terms of the variables $A_{i}^{a}$ and $E^{bj}$, is
finally obtained and compared with the standard YM Hamiltonian. The
conditions under which both theories are equivalent are established, which
requires understanding the time evolution of the Gauss functions $\Omega ^{a}
$ under the SL-NANM dynamics. A similar analysis can be carried out for each
of the remaining cases, corresponding to the time-like and light-like NANM.
Section III presents a substantial conceptual and practical improvement over
the previous individual calculations to study the relation between the NANM
and YM theories. We start from a general parametrization that solves the
constraint (\ref{NANMC}) for arbitrary values of $n^{2}$, in terms of $3N$
d.o.f. $\,\Phi _{A}^{a},A=1,2,3$, and repeat the canonical analysis, paying
attention to the relations between the standard canonical variables of the YM
theory $(A_{i}^{a},E^{bj})$ and those of the NANM $(\Phi _{A}^{a},\Pi
_{B}^{b})$. The calculations are enormously simplified after one realizes
that the transformation $(\Phi _{A}^{a},\Pi _{B}^{b})\rightarrow
(A_{i}^{a},E^{bj})$ is a canonical transformation, once the $E^{bj}$ are
recognized as the momenta canonically conjugate to the $A_{i}^{a}$ via the
kinetic part of the NANM Hamiltonian action. Another useful property of the
new parametrization is that it exhibits the NANM as a regular theory (i.e.,
no constraints appear in the Hamiltonian analysis). This is proved in 
Appendix C. The canonical NANM Hamiltonian, rewritten in terms of the YM
variables $(A_{i}^{a},E^{bj})$, is finally obtained and the conditions under
which it reduces to the YM Hamiltonian are determined. Again, this requires
the calculation of the time evolution of the Gauss functions $\Omega ^{a}$
under the NANM dynamics. Finally, we close with a summary and some comments
in Sec. IV. Appendix A serves to establish notation and briefly reviews
the canonical version of the SO$(N)$ YM theory, which we use as a benchmark
to identify the conditions under which it is equivalent to the different
realizations of the NANM.

\section{The space-like case of the non-Abelian Nambu model (SL-NANM)}

\label{sectionSL}

Before considering the specific case of the SL-NANM, let us recall some
general properties of the NANM. The\ Lagrangian is 
\begin{equation}
\mathcal{L}_{NANM}(A_{\mu }^{a})=-\frac{1}{4}F_{\mu \nu }^{a}F^{a\mu \nu
}-J^{a\mu }A_{\mu }^{a}\,+\lambda \left( A_{\mu }^{a}A^{a\mu
}-n^{2}M^{2}\right) ,\,\ \ \
\,\,\,\,\,\,\,\,\,\,\,\,\,M^{2}>0,\;\;\;\;\;a=1,...,N,  \label{LNANM}
\end{equation}%
with the notation and conventions introduced in Appendix A. Here, $\lambda $  is a Lagrange multiplier and the vector\ $n^{\mu }\;$is such
that $n^{2}=\pm 1,0.$

The general procedure by which we will analyze each model is to explicitly
solve the condition 
\begin{equation}
A_{\mu }^{a}A^{a\mu }=n^{2}M^{2}  \label{NACOND}
\end{equation}%
and substitute the adequate parametrizations directly into the Lagrangian $%
\mathcal{L}_{\rm NANM}(A_{\mu }^{a})$, defining in this way the canonical
degrees of freedom of the model. Subsequently, we obtain the corresponding
Hamiltonian and determine its relation to the YM Hamiltonian (\ref%
{hamiltonian_standar}) together with the canonical algebra (\ref{FINALDB}).
As expected, the equations of motion of the NANM will not be those of the \
Yang-Mills theory arising from $\mathcal{L}(\mathbf{A}^{\mu })\;$in Eq.\ (%
\ref{LDENS}). This property will be explicitly shown in the remaining sections
of the paper. In this way, the conservation of the current $J_{\mu }^{a}\;$%
does not follow as a consistency condition from the equations of motion in
the NANM, as happens in the YM case. We will show that the canonical
structure of the NANM will induce the standard-algebra YM [(\ref{FINALDB})]
together with a Hamiltonian that differs from Eq. (\ref{hamiltonian_standar}) by
the property that the Gauss laws do not appear as constraints. Nevertheless,
the dynamics of the NANM guarantees their validity for all time, once they
are imposed as initial conditions.

The standard solutions of the condition $A_{\mu }^{a}A^{a\mu }=n^{2}M^{2}$%
, arising from the different choices of $n^{2}$, are 
\begin{eqnarray}
n^{2} &>&0:A_{0}^{1}=\sqrt{M^{2}+A_{i}^{a}A_{i}^{a}-A_{0}^{\bar{b}}A_{0}^{%
\bar{b}}}\,\,,\,\,\,\,\,\,\,\,\,\,\bar{b}=2,3,..,N,  \label{NANM_TL} \\
n^{2} &<&0:\;A_{3}^{1}=\sqrt{%
M^{2}+(A_{0}^{a})^{2}-(A_{1}^{a})^{2}-(A_{2}^{a})^{2}-(A_{3}^{\bar{b}})^{2}}
\label{NANM_SL} \\
n^{2} &=&0:\;A_{0}^{a}=B^{a}\left( 1+\frac{A_{\bar{\imath}}^{b}A_{\bar{\imath%
}}^{b}}{4B^{2}}\right) ,\,\,\,\,\,\,\,\,\,\,\,\,\,\,\,A_{3}^{a}=B^{a}\left(
1-\frac{A_{\bar{\imath}}^{b}A_{\bar{\imath}}^{b}}{4B^{2}}\right)
\,\,\,\,\,\,\,\,\,\,\,\bar{\imath}=1,2,\;\;\;\,  \label{NANM_LL}
\end{eqnarray}%
which define the NANM in its time-like (TL-NANM), space-like (SL-NANM) and
light-like (LL-NANM) representations. In the time-like and space-like cases
we start with $4N-1$ d.o.f. per point, while in the light-like case this number
is $3N$. Next we concentrate on the SL-NANM.

\subsection{The equations of motion in the SL-NANM}

We start with the extension to the non-Abelian case of the parametrization $%
A_{3}=\sqrt{M^{2}+A_{0}^{2}-A_{1}^{2}-A_{2}^{2}}$, which is frequently used
in the Abelian case to exhibit the remaining SO$(2,1)$ symmetry after the
spontaneous Lorentz symmetry breaking. The Lagrangian constraint (\ref%
{NACOND}) is now solved for $A_{\mu =3}^{a=1}$. In this way we start from $%
4N-1$ d.o.f.\ in coordinate space, which we denote in the following way: $%
\;\;A_{0}^{a},\;A_{\bar{k}}^{1},\;\;A_{k}^{\bar{a}}$, with $a=1,2,...,N;\;%
\bar{a}=2,3,...,N,\;k=1,2,3;\;\bar{k}=1,2$. The numbers of each of the
corresponding fields are $N,\;2$, $3(N-1)$, respectively. With this
notation, we write 
\begin{equation}
A_{3}^{1}=\sqrt{M^{2}+(A_{0}^{a})^{2}-(A_{\bar{k}}^{1})^{2}-(A_{k}^{\bar{a}%
})^{2}},  \label{constr_space}
\end{equation}%
which exhibits the remaining symmetry group SO$(N, 3N-1)$. As a matter of
notation, superscript indices label a group index, while subscript indices
refers to a space-time index.

In the notation of Appendix A, the Lagrangian density (\ref{LNANM})
takes the form 
\begin{equation}
\mathcal{L}(A_{0}^{a},A_{\bar{\imath}}^{1},A_{k}^{\bar{a}})=\frac{1}{2}%
\left( \frac{{}}{{}}(E_{i}^{\bar{a}})^{2}+(E_{\bar{\imath}%
}^{1})^{2}+(E_{3}^{1})^{2}-B_{k}^{a}B_{k}^{a}\right) -J_{0}^{a}A^{a0}+J_{%
\bar{k}}^{1}A_{\bar{k}}^{1}+J_{i}^{\bar{a}}A_{i}^{\bar{a}%
}+J_{3}^{1}A_{3}^{1}.  \label{LSPACELIKE}
\end{equation}%
The equations of motion are%
\begin{equation}
\mathcal{E}^{ia}-\mathcal{E}^{31}\frac{A_{i}^{a}}{A_{3}^{1}}=0\,,
\label{SL_LAG1}
\end{equation}%
\begin{equation}
\mathcal{E}^{0a}+\mathcal{E}^{31}\frac{A_{0}^{a}}{A_{3}^{1}}=0,
\label{SL_LAG2}
\end{equation}%
with the notation 
\begin{equation}
\mathcal{E}^{\nu a}=\left( D_{\mu }F^{\mu \nu }-J^{\nu }\right)^{a}.
\label{EQMOTYM}
\end{equation}%
The numbers of equations in  Eq. (\ref{SL_LAG1}) is only $(3N-1)$ because the simultaneous
choice $i=3$ and $a=1$ does not appear, as $A_{3}^{1}$ is a function of the
dynamical variables.

Since $A_{3}^{1}$ is just a shorthand for (\ref{constr_space}), it follows
that 
\begin{equation}
\dot{A}_{3}^{1}=\frac{A_{0}^{a}}{A_{3}^{1}}\dot{A}_{0}^{a}-\frac{A_{\bar{k}%
}^{1}}{A_{3}^{1}}\dot{A}_{\bar{k}}^{1}-\frac{A_{i}^{\bar{a}}}{A_{3}^{1}}\dot{%
A}_{i}^{\bar{a}}.  \label{A31}
\end{equation}%
The fields $E_{3}^{1}$ and $E_{\bar{\imath}}^{\bar{a}}$ are given by 
\begin{eqnarray}
E_{3}^{1} &=&\dot{A}_{3}^{1}-D_{3}A_{0}^{1}=\left( \frac{A_{0}^{a}}{A_{3}^{1}%
}\dot{A}_{0}^{a}-\frac{A_{\bar{k}}^{1}}{A_{3}^{1}}\dot{A}_{\bar{k}}^{1}-%
\frac{A_{i}^{\bar{a}}}{A_{3}^{1}}\dot{A}_{i}^{\bar{a}}\right)
-D_{3}A_{0}^{1},  \nonumber \\
E_{\bar{k}}^{1} &=&\dot{A}_{\bar{k}}^{1}-D_{\bar{k}}A_{0}^{1},\;\;\;\;\;\;\;%
\;E_{i}^{\bar{a}}=\dot{A}_{i}^{\bar{a}}-D_{i}A_{0}^{\bar{a}},
\label{campos_electricos2}
\end{eqnarray}%
which, for the moment, constitute a compact way of identifying the
velocities $\dot{A}_{0}^{a}\;,\;\dot{A}_{i}^{\bar{a}}$, and $\dot{A}_{\bar{%
\imath}}^{1}$.

\subsection{The Hamiltonian density of the SL-NANM}

The canonically conjugate momenta are 
\begin{eqnarray}
\Pi _{0}^{a} &=&\frac{\partial \mathcal{L}}{\partial \dot{A}_{0}^{a}}%
=E_{3}^{1}\frac{A_{0}^{a}}{A_{3}^{1}},  \label{PI0} \\
\Pi ^{1\bar{k}} &=&\frac{\partial \mathcal{L}}{\partial \dot{A}_{\bar{k}}^{1}%
}=E_{\bar{k}}^{1}-E_{3}^{1}\frac{A_{\bar{k}}^{1}}{A_{3}^{1}},  \label{PI1BAR}
\\
\Pi ^{\bar{a}i} &=&\frac{\partial \mathcal{L}}{\partial \dot{A}_{i}^{\bar{a}}%
}=E_{i}^{\bar{a}}-E_{3}^{1}\frac{A_{i}^{\bar{a}}}{A_{3}^{1}},  \label{PIBARI}
\end{eqnarray}%
with nonzero Poisson brackets (PBs)%
\begin{equation}
\,\,\,\,\,\,\,\,\,\,\,\{A_{0}^{a}(\mathbf{x},t),\Pi _{0}^{b}(\mathbf{y}%
,t)\}=\delta ^{3}(\mathbf{x}-\mathbf{y})\delta ^{ab}\,\,\,,\;\;\;\{A_{\bar{%
\imath}}^{1}(\mathbf{x},t),\Pi ^{1\bar{k}}(\mathbf{y},t)\}=\delta _{\bar{%
\imath}}^{\bar{k}}\delta ^{3}(\mathbf{x}-\mathbf{y}),\;\;\;\;\{A_{i}^{\bar{a}%
}(\mathbf{x},t),\Pi ^{\bar{b}j}(\mathbf{y},t)\}=\delta _{i}^{j}\delta ^{%
\bar{a}\bar{b}}\delta ^{3}(\mathbf{x}-\mathbf{y}).  \label{CA_SLNANM}
\end{equation}%
Next we solve for the velocities. From Eqs. (\ref{PI0}), (\ref{PI1BAR}) and (%
\ref{PIBARI}) we (respectively) obtain 
\begin{eqnarray}
\left( \frac{A_{0}^{a}}{A_{3}^{1}}\dot{A}_{0}^{a}-\frac{A_{\bar{k}}^{1}}{%
A_{3}^{1}}\dot{A}_{\bar{k}}^{1}-\frac{A_{i}^{\bar{a}}}{A_{3}^{1}}\dot{A}%
_{i}^{\bar{a}}\right) -D_{3}A_{0}^{1} &=&\frac{\Pi _{0}^{1}}{A_{0}^{1}}%
A_{3}^{1},  \label{VELSL1} \\
\dot{A}_{\bar{k}}^{1}-D_{\bar{k}}A_{0}^{1}-\frac{\Pi _{0}^{1}}{A_{0}^{1}}A_{%
\bar{k}}^{1} &=&\Pi ^{1\bar{k}},  \label{VELSL12} \\
\dot{A}_{i}^{\bar{a}}-D_{i}A_{0}^{\bar{a}}-\frac{\Pi _{0}^{1}}{A_{0}^{1}}%
A_{i}^{\bar{a}} &=&\Pi ^{\bar{a}i}.  \label{VELSL2}
\end{eqnarray}%
From (\ref{VELSL12}) and (\ref{VELSL2}) we can solve for 
\begin{eqnarray}
\dot{A}_{\bar{k}}^{1} &=&\Pi ^{1\bar{k}}+D_{\bar{k}}A_{0}^{1}+\frac{\Pi
_{0}^{1}}{A_{0}^{1}}A_{\bar{k}}^{1}, \\
\dot{A}_{i}^{\bar{a}} &=&\Pi ^{\bar{a}i}+D_{i}A_{0}^{\bar{a}}+\frac{\Pi
_{0}^{1}}{A_{0}^{1}}A_{i}^{\bar{a}}.
\end{eqnarray}%
We can substitute these velocities into Eq. (\ref{VELSL1}), but we cannot
solve for all the $\dot{A}_{0}^{a}$ which enter into the sum $A_{0}^{a}\dot{A}%
_{0}^{a}$. At most we could solve for one velocity, say $\dot{A}%
_{0}^{1}$, in terms of the remaining $(N-1)\;\dot{A}_{0}^{\bar{a}}$. This is
consistent with the existence of $\left( N-1\right) $ primary constraints $%
\Phi _{1}^{\bar{a}},\;$which we choose as 
\begin{equation}
\Phi _{1}^{\bar{a}}=\Pi _{0}^{\bar{a}}-\frac{\Pi _{0}^{1}}{A_{0}^{1}}A_{0}^{%
\bar{a}},  \label{PC_SL}
\end{equation}%
arising from Eq. (\ref{PI0}). It is more convenient to consider the solved
velocity as$\;\dot{A}_{3}^{1}$, encoded in the definition of$\;E_{3}^{1}\;$%
and written in terms of the canonical variables as 
\begin{equation}
E_{3}^{1}=\Pi _{0}^{1}\frac{A_{3}^{1}}{A_{0}^{1}}, \label{E13}
\end{equation}%
via the remaining relation (\ref{PI0}) corresponding to $a=1.$

From Eqs. (\ref{PI0}), (\ref{PI1BAR}) and (\ref{PIBARI}) we can express the
electric fields in terms of the canonical momenta as 
\begin{equation}
E_{3}^{1}=A_{3}^{1}\left( \frac{\Pi _{0}^{1}}{A_{0}^{1}}\right)
,\;\;\;\;\;\;E_{\bar{\imath}}^{1}=\Pi ^{1\bar{\imath}}+\left( \frac{\Pi
_{0}^{1}}{A_{0}^{1}}\right) A_{\bar{\imath}}^{1}\;,\;\;\;\;\;E_{i}^{\bar{a}%
}=\Pi ^{\bar{a}i}+\left( \frac{\Pi _{0}^{1}}{A_{0}^{1}}\right) A_{i}^{\bar{a}%
},  \label{1}
\end{equation}%
where we have used Eq. (\ref{E13}).

In Appendix B we show that the above definitions of the electric fields $%
E_{i}^{a}$ in terms of the canonical momenta $\Pi ^{ai}$, together with the
canonical algebra (\ref{CA_SLNANM}) lead to the following PB relations 
\begin{equation}
\;\;\;\;\{A_{i}^{a}(\mathbf{x},t),A_{j}^{b}(\mathbf{y},t)\}=0,\,\,\,\,\,\,\,%
\,\,\,\,\,\{E^{ai}(\mathbf{x},t),E^{bj}(\mathbf{y},t)\}=0,\;\;\{A_{i}^{a}(%
\mathbf{x},t),E^{bj}(\mathbf{y},t)\}=- \delta _{i}^{j}\delta ^{ab}\delta (%
\mathbf{x}-\mathbf{y}),  \label{AE_SL_ALGEBRA}
\end{equation}%
which reproduces the YM algebra (\ref{FINALDB}). In the following we will
also need the PB of the variables $A_{0}^{a},\;\Pi _{0}^{a}$ with $A_{i}^{a}$
and $E_{i}^{a}$, which are summarized in Eqs. (\ref{PBA0E1}) and (\ref%
{PBPI0R}) of Appendix B.

The next step is to obtain the Hamiltonian density 
\begin{equation}
\mathcal{H}=\Pi _{0}^{a}\dot{A}_{0}^{a}+\Pi ^{1\bar{\imath}}\dot{A}_{\bar{%
\imath}}^{1}+\Pi ^{\bar{a}k}\dot{A}_{k}^{\bar{a}}-L(A_{0}^{a},A_{\bar{\imath}%
}^{\bar{a}}),  \label{HSPACELIKE}
\end{equation}%
where $\mathcal{L}(A_{0}^{a},A_{\bar{\imath}}^{1},A_{k}^{\bar{a}})$ is given
in Eq. (\ref{LSPACELIKE}). The main goal is to rewrite this Hamiltonian in
terms of the fields $E_{i}^{a}$ and $A_{j}^{c}$ in order to compare with the YM
result (\ref{hamiltonian_standar}). From Eqs. (\ref{PI0}), (\ref{PI1BAR})
and (\ref{PIBARI}), we substitute the momenta into the above equation
obtaining 
\begin{eqnarray}
\mathcal{H} &=&E_{3}^{1}\frac{A_{0}^{a}}{A_{3}^{1}}\dot{A}_{0}^{a}+\left( E_{%
\bar{\imath}}^{1}-E_{3}^{1}\frac{A_{\bar{\imath}}^{1}}{A_{3}^{1}}\right) 
\dot{A}_{\bar{\imath}}^{1}+\left( E_{k}^{\bar{a}}-E_{3}^{1}\frac{A_{k}^{%
\bar{a}}}{A_{3}^{1}}\right) \dot{A}_{k}^{\bar{a}}-L(A_{0}^{a},A_{\bar{\imath}%
}^{\bar{a}}), \\
&=&E_{3}^{1}\left( \frac{A_{0}^{a}}{A_{3}^{1}}\dot{A}_{0}^{a}-\frac{A_{\bar{%
\imath}}^{1}}{A_{3}^{1}}\dot{A}_{\bar{\imath}}^{1}-\frac{A_{k}^{\bar{a}}}{%
A_{3}^{1}}\dot{A}_{k}^{\bar{a}}\right) +E_{\bar{\imath}}^{1}\dot{A}^{1\bar{%
\imath}}+E_{k}^{\bar{a}}\dot{A}^{\bar{a}k}-L(A_{0}^{a},A_{\bar{\imath}}^{%
\bar{a}}),
\end{eqnarray}%
where we subsequently replace the velocities $\dot{A}_{\bar{\imath}}^{1}\;$%
and $\dot{A}_{k}^{\bar{a}}$ in terms of the electric fields, using Eqs. (\ref%
{campos_electricos2}). This yields 
\begin{eqnarray}
\mathcal{H} &=&E_{\bar{k}}^{1}\left( E_{\bar{k}}^{1}+D_{\bar{k}%
}A_{0}^{1}\right) +E_{i}^{\bar{a}}\left( E_{i}^{\bar{a}}+D_{i}A_{0}^{\bar{a}%
}\right) +E_{3}^{1}\left( E_{3}^{1}+D_{3}A_{0}^{1}\right)  \nonumber \\
&&-\frac{1}{2}\left( \frac{{}}{{}}(E_{i}^{\bar{a}})^{2}+(E_{\bar{\imath}%
}^{1})^{2}+(E_{3}^{1})^{2}-B_{k}^{a}B_{k}^{a}\right) +J_{0}^{a}A^{a0}-J_{%
\bar{k}}^{1}A_{\bar{k}}^{1}-J_{i}^{\bar{a}}A_{i}^{\bar{a}}-J_{3}^{1}A_{3}^{1}
\end{eqnarray}%
which can be further rearranged as 
\begin{eqnarray}
\mathcal{H} &=&\frac{1}{2}E_{i}^{\bar{a}}E_{i}^{\bar{a}}+\frac{1}{2}E_{\bar{%
\imath}}^{1}E_{\bar{\imath}}^{1}+\frac{1}{2}E_{3}^{1}E_{3}^{1}+\frac{1}{2}%
B_{k}^{a}B_{k}^{a}+J_{0}^{a}A^{a0}-J_{\bar{k}}^{1}A_{\bar{k}}^{1}-J_{i}^{%
\bar{a}}A_{i}^{\bar{a}}-J_{3}^{1}A_{3}^{1}  \nonumber \\
&&+E_{\bar{k}}^{1}D_{\bar{k}}A_{0}^{1}+E_{i}^{\bar{a}}D_{i}A_{0}^{\bar{a}%
}+E_{3}^{1}D_{3}A_{0}^{1},  \nonumber \\
\mathcal{H} &=&\frac{1}{2}E_{i}^{a}E_{i}^{a}+\frac{1}{2}%
B_{k}^{a}B_{k}^{a}-J_{i}^{a}A_{i}^{a}-A_{0}^{a}\left(
D_{i}E_{i}^{a}-J_{0}^{a}\right) .  \label{HAMFINSL}
\end{eqnarray}%
Here we have integrated by parts the term $E_{i}^{a}\left(
D_{i}A_{0}^{a}\right) $ in the Hamiltonian. The form of the above
Hamiltonian density, together with the PBs in Eq. (\ref{AE_SL_ALGEBRA}), is
similar to that of the SO$(N)$\ YM theory (\ref{hamiltonian_standar}),
except for the following facts: (i) the coordinates $A_{0}^{a}$ are
dynamical instead of being Lagrange multipliers, (ii) the Gauss functions,
defined as 
\begin{equation}
\Omega ^{a}=\left( D_{i}E_{i}^{a}-J_{0}^{a}\right) =\mathcal{E}^{0a},
\label{NOTGL}
\end{equation}
are not constraints in the SL-NANM and (iii) we have the additional primary
constraints (\ref{PC_SL}).

Then we need to continue the Hamiltonian analysis by applying Dirac's
procedure starting from the extended Hamiltonian density 
\begin{equation}
\mathcal{H}_{E}=\frac{1}{2}E_{i}^{a}E_{i}^{a}+\frac{1}{2}%
B_{k}^{a}B_{k}^{a}-J_{i}^{a}A_{i}^{a}-A_{0}^{a}\Omega ^{a}+\mu ^{\bar{a}%
}\left( \Pi _{0}^{\bar{a}}-\frac{\Pi _{0}^{1}}{A_{0}^{1}}A_{0}^{\bar{a}%
}\right) .  \label{HEXTENDED}
\end{equation}%
The evolution of the primary constraints yields%
\begin{eqnarray}
\dot{\Phi}_{1}^{\bar{a}} &=&\left\{ \Phi _{1}^{\bar{a}},\;\int
d^{3}y\;\left( \frac{1}{2}E_{k}^{b}E_{k}^{b}+\frac{1}{2}%
B_{k}^{a}B_{k}^{a}-J_{i}^{a}A_{i}^{a}-A_{0}^{a}\left(
D_{i}E_{i}^{a}-J_{0}^{a}\right) \right) \right\} ,  \nonumber \\
\dot{\Phi}_{1}^{\bar{a}} &=&\left\{ \Phi _{1}^{\bar{a}},\;\int
d^{3}y\;\left( -A_{0}^{1}\Omega ^{1}-A_{0}^{\bar{b}}\Omega ^{\bar{b}}\right)
\right\} ,
\end{eqnarray}
the calculation of which  requires the following PBs calculated in Appendix B 
\begin{eqnarray}
\left\{ \Phi _{1}^{\bar{a}},\;\;A_{k}^{a}\right\} &=&0,\;\;\;\;\left\{ \Phi
_{1}^{\bar{a}},\;E_{k}^{b}\;\right\} =0,\;\;\;\left\{ \Phi _{1}^{\bar{a}%
},\;B_{k}^{b}\;\right\} =0,  \nonumber \\
\left\{ \Phi _{1}^{\bar{a}},\;A_{0}^{1}\;\right\} &=&\frac{A_{0}^{\bar{a}}}{%
A_{0}^{1}},\;\;\;\;\left\{ \Phi _{1}^{\bar{a}},\;A_{0}^{\bar{b}}\;\right\}
=-\delta ^{\bar{a}\bar{b}}.
\end{eqnarray}%
In this way, the only contribution arises from the terms proportional to $%
A_{0}^{a}$ in the Hamiltonian density. The result 
\begin{equation}
\dot{\Phi}_{1}^{\bar{a}}=-\;\frac{A_{0}^{\bar{a}}}{A_{0}^{1}}\Omega
^{1}+\delta ^{\bar{a}\bar{b}}\Omega ^{\bar{b}}
\end{equation}%
produces secondary constraints, which we write as 
\begin{equation}
\Phi _{2}^{\bar{a}}=A_{0}^{\bar{a}}-\frac{A_{0}^{1}}{\Omega ^{1}}\Omega ^{%
\bar{a}}.
\end{equation}%
Next we calculate the time evolution of $\Phi _{2}^{\bar{a}}$ using the
relations 
\begin{equation}
\left\{ \Omega ^{\bar{a}},\;\Phi _{1}^{\bar{c}}\right\} =0,\;\;\left\{ \Phi
_{1}^{\bar{a}},\;A_{0}^{\bar{b}}\;\right\} =-\delta ^{\bar{a}\bar{b}%
}, \;\;\;\;\left\{ \frac{A_{0}^{1}}{\Omega ^{1}}\Omega ^{\bar{a}},\;\Phi _{1}^{%
\bar{c}}\right\} =-\frac{\Omega ^{\bar{a}}}{\Omega ^{1}}\frac{A_{0}^{\bar{c}}%
}{A_{0}^{1}},
\end{equation}%
which are included in Appendix B. We obtain 
\[
\dot{\Phi}_{2}^{\bar{a}}=W^{\bar{a}}+\mu ^{\bar{a}}+\frac{A_{0}^{\bar{a}%
}A_{0}^{\bar{c}}}{\left( A_{0}^{1}\right) ^{2}}\mu ^{\bar{c}}. 
\]%
In fact, we can solve 
\begin{equation}
\left( \delta ^{\bar{a}\bar{c}}+\frac{A_{0}^{\bar{a}}A_{0}^{\bar{c}}}{\left(
A_{0}^{1}\right) ^{2}}\right) \mu ^{\bar{c}}\;=-W^{\bar{a}}
\end{equation}%
for the arbitrary functions $\mu ^{\bar{a}}$, concluding that 
\begin{equation}
\mu ^{\bar{a}}=-\left( \delta ^{\bar{a}\bar{c}}-\frac{1}{\left(
A_{0}^{b}A_{0}^{b}\right) }A_{0}^{\bar{a}}A_{0}^{\bar{c}}\right) W^{\bar{c}}.
\end{equation}%
In this way the Dirac method stops and we are left with $2(N-1)$ constraints 
\begin{equation}
\Phi _{1}^{\bar{a}}=\Pi _{0}^{\bar{a}}-\frac{\Pi _{0}^{1}}{A_{0}^{1}}A_{0}^{%
\bar{a}},\qquad \Phi _{2}^{\bar{a}}=\;A_{0}^{\bar{a}}-\frac{A_{0}^{1}}{%
\Omega ^{1}}\Omega ^{\bar{a}},  \label{SCC_NANM}
\end{equation}%
which are second class. Thus the number of d.o.f. per point of the SL-NANM is 
\begin{equation}
\#d.o.f.=\frac{1}{2}(2(4N-1)-2(N-1))=3N,
\end{equation}%
which does not correspond to the number of d.o.f. of the SO$(N)$ Yang-Mills
theory.

The next step is to set  the constraints (\ref%
{SCC_NANM}) strongly equal to zero, in order to eliminate the variables $A_{0}^{\bar{a}}$ and  $\Pi
_{0}^{\bar{c}}$, and to subsequently introduce the corresponding Dirac
brackets among the remaining variables. To this end we require the matrix
constructed with the PB of the constraints 
\begin{equation}
M=\left[ 
\begin{array}{cc}
\left[ R^{\bar{a}\bar{b}}\right] & \left[ T^{\bar{a}\bar{b}}\right] \\ 
-\left[ T^{\bar{b}\bar{a}}\right] & \;\left[ S^{\bar{a}\bar{b}}\right]%
\end{array}%
\right] =\left[ 
\begin{array}{cc}
R & T \\ 
-T^{T} & \;S%
\end{array}%
\right] ,  \label{SL_PP_MATRIX}
\end{equation}%
where 
\begin{equation}
R^{\bar{a}\bar{b}}=\left\{ \Phi _{1}^{\bar{a}},\;\;\Phi _{1}^{\bar{b}%
}\right\} ,\;\;\;T^{\bar{a}\bar{b}}=\left\{ \Phi _{1}^{\bar{a}},\;\;\Phi
_{2}^{\bar{b}}\right\} ,\;\;\;\;S^{\bar{a}\bar{b}}=\left\{ \Phi _{2}^{\bar{a}%
},\;\;\Phi _{2}^{\bar{b}}\right\} \;.\;  \label{SL_PP_ENTRIES}
\end{equation}%
The required calculations produce 
\begin{eqnarray}
R^{\bar{a}\bar{b}} &=&0,\qquad T^{\bar{a}\bar{b}}=\left\{ \Phi _{1}^{\bar{a}%
},\;\Phi _{2}^{\bar{b}}\right\} =-\left( \delta ^{\bar{a}\bar{b}}+\frac{%
A_{0}^{\bar{a}}}{A_{0}^{1}}\frac{A_{0}^{\bar{b}}}{A_{0}^{1}}\right) =T^{\bar{%
b}\bar{a}},  \nonumber \\
S^{\bar{a}\bar{b}} &=&\left( \frac{A_{0}^{1}}{\Omega ^{1}}\right) ^{2}\left(
C^{\bar{a}\bar{b}m}+C^{1\bar{a}m}\frac{\Omega ^{\bar{b}}}{\Omega ^{1}}-C^{1%
\bar{b}m}\frac{\Omega ^{\bar{a}}}{\Omega ^{1}}\right) \left(
D_{i}E_{i}^{m}\right) .
\end{eqnarray}%
according to the results in Appendix B.

The matrix $T$ is invertible, yielding 
\begin{equation}
\left( T^{-1}\right)^{\bar{a}\bar{b}}=-\left( \delta ^{\bar{a}\bar{b}}-\frac{%
A_{0}^{\bar{a}}A_{0}^{\bar{b}}}{A_{0}^{m}A_{0}^{m}} \right),
\end{equation}%
in such a way that 
\begin{equation}
\;\;M^{-1}=\left[ 
\begin{array}{cc}
T^{-1}ST^{-1} & -T^{-1} \\ 
T^{-1} & 0%
\end{array}%
\right] .
\end{equation}%
The Dirac bracket is 
\begin{eqnarray}
\{A(x),B(y)\}^{\ast } &=&\{A(x),B(y)\}-\{A,\phi _{1}^{\bar{a}%
}\}(T^{-1}ST^{-1})^{\bar{a}\bar{b}}\{\phi _{1}^{\bar{b}},B\}  \nonumber \\
&&+\{A,\phi _{1}^{\bar{a}}\}(T^{-1})^{\bar{a}\bar{b}}\{\phi _{2}^{\bar{b}%
},B\}-\{A,\phi _{2}^{\bar{a}}\}(T^{-1})^{\bar{a}\bar{b}}\{\phi _{1}^{\bar{b}%
},B\},  \label{DB_TLNANM1}
\end{eqnarray}%
which leads to the result 
\begin{equation}
\{A(x),B(y)\}^{\ast }=\{A(x),B(y)\},  \label{DBEPBSL}
\end{equation}%
for the YM variables $A_{i}^{a}$ and $E^{bj}$. The above conclusion arises
from the fact that each of the additional PBs in Eq. (\ref{DB_TLNANM1})
includes a contribution from $\phi _{1}^{\bar{a}}$, which has zero PB with
those variables, according to Eq. (\ref{CPHI11}). In other words, we
recover the algebra 
\begin{equation}
\{A_{i}^{a}(\mathbf{x},t),A_{j}^{b}(\mathbf{y},t)\}^{\ast }=0=\{E^{ai}(%
\mathbf{x},t),E^{bj}(\mathbf{y},t)\}^{\ast },\;\;\;\;\{A_{i}^{a}(\mathbf{x}%
,t),E^{bj}(\mathbf{y},t)\}^{\ast }=-\delta _{i}^{j}\delta ^{ab}\delta (%
\mathbf{x}-\mathbf{y}),
\end{equation}%
corresponding to the YM theory given in Eq. (\ref{FINALDB}) of Appendix
A. Having set  the constraints (\ref{SCC_NANM}) strongly equal to zero, the extended
Hamiltonian (\ref{HEXTENDED}) now reduces to 
\begin{equation}
\mathcal{H}_{E}=\frac{1}{2}E_{i}^{a}E_{i}^{a}+\frac{1}{2}%
B_{k}^{a}B_{k}^{a}-J_{i}^{a}A_{i}^{a}-A_{0}^{a}\Omega ^{a}.
\end{equation}%
but we are still\thinspace\ missing the Gauss laws $\Omega ^{a}=0$, because
the $A_{0}^{a}\;$are dynamical degrees of freedom.

\subsection{The evolution of the Gauss functions $\Omega ^{a}\;$in the
SL-NANM}

Next we study the time evolution of the Gauss functions, starting from the
Hamiltonian density (\ref{HAMFINSL}). A direct use of (\ref{FINALDB}) leads
to 
\begin{equation}
\dot{\Omega}^{a}=-gC^{abc}A_{0}^{b}\Omega ^{c}-D_{\mu }J^{\mu a}-\int
d^{3}y\left\{ \Omega ^{a}(x),\;A_{0}^{b}(y)\right\} \Omega ^{b}(y).
\label{TEVOLGF}
\end{equation}%
where $\Omega ^{a}(x)\rightarrow \left( \delta ^{ac}\partial
_{kx}+gC^{abc}A_{k}^{b}\right) E_{k}^{c}\;$inside a PB because $J^{0a}$ has
been considered as an external current. Since $\left\{
A_{0}^{a},\;A_{k}^{b}\right\} =0$, we only need the brackets $\left\{
E_{k}^{c}(x),\;A_{0}^{b}(y)\right\} $. Using Eq. (\ref{PBA0E1}) we obtain%
\begin{equation}
\dot{\Omega}^{a}=-gC^{abc}A_{0}^{b}\Omega ^{c}-D_{\mu }J^{\mu a}+D_{k}\left(
\left( \frac{A_{k}^{a}}{A_{0}^{1}}\right) \Omega ^{1}\right) .
\label{RESULTSLNANM}
\end{equation}%
The above equations guarantee that by (i) imposing current conservation $%
D_{\mu }J^{\mu a}=0$ at some initial time $t=0$ and (ii) demanding that the
Gauss laws $\Omega ^{a}=0$  hold at $t=0$, we obtain $\partial _{0}\Omega
^{a}=0$ $(a=1,2,...,N)$ as well at $t=0$. This is enough to prove that with
these two initial conditions, the Gauss laws will hold for all time. In this
way we can recover the SO$(N)$ Yang-Mills theory by imposing the Gauss laws
as Hamiltonian constraints, with arbitrary functions $N^{a}$ adding $%
-N^{a}\Omega ^{a}$ to $\mathcal{H}_{E}$ and redefining $A_{0}^{a}+N^{a}=%
\Theta ^{a}$. This leads to 
\begin{equation}
\mathcal{H}_{E}=\frac{1}{2}(\mathbf{E}^{2}+\mathbf{B}^{2})-\Theta ^{a}\Omega
^{a}+J_{i}^{a}A^{ia},  \label{HAMFIN_SLNANM}
\end{equation}%
where the $\Theta ^{a}$ are now arbitrary functions, and thus we get  back to the YM
Hamiltonian density (\ref{hamiltonian_standar}). The subsequent emergence of the SO$(N)$ YM theory guarantees current conservation for all times, as a
consequence of the equations of motion.

From the perspective of the GB modes, the situation in the SL-NANM is as
follows. We have started from a theory invariant under SO$(N,3N)$ defined
by Eqs. (\ref{YML}) and (\ref{NANMC}). Solving the constraint (\ref{NANMC})
in terms of $A_{3}^{1}$ means that we have the nonzero vacuum expectation
value $\langle A_{3}^{1}\rangle =M$, which breaks the symmetry down to 
SO$(N,3N-1)$,  with the appearance of $(4N(4N-1)/2-(4N-1)(4N-2)/2)=\allowbreak
4N-1$ GBs. Nevertheless, the SL-NANM phase space still contains $2(N-1)$
second-class constraints [Eq. (\ref{SCC_NANM})] which can be imposed strongly,
yielding a reduced phase space with $\frac{1}{2}(2(4N-1)-2(N-1)=\allowbreak
3N$ coordinates per point. A similar analysis yields $3N$ as the number of
independent GB modes in each realization of the NANM. This conclusion is
consistent with Ref. \cite{JLCH2}, where the final independent GB modes
are denoted by $a_{\mu ^{\prime }}^{i}$ $(i=1,..,N,\;\;\mu ^{\prime
}=1,2,3)\;$for the time-like case and $a_{\mu ^{\prime \prime }}^{i}$ $%
(i=1,..,N,\;\;\mu ^{\prime \prime }=0,1,2)$ for the space-like case. Once
more, we verify that in order to regain the $2N$ independent vector bosons
of the YM theory, we still have to impose the $N$ Gauss laws $\Omega
^{a}=0\;$as first-class constraints. In this way we are left with $\frac{1}{2%
}(6N-2N)=2N$ coordinates per point.

To summarize, the emergence of the SO$(N)$ YM theory from the SL-NANM can be
established only after imposing both current conservation and the
Gauss laws as initial conditions.

\section{ A unified description of the non-Abelian Nambu models}

A procedure similar to that presented in the previous section for the
SL-NANM can be repeated for the TL-NANM and the LL-NANM, with identical
results. The standard variables $A_{i}^{a}$, $E_{j}^{b}$ of the YM theory
can be expressed in terms of the canonical variables of each version of the
NANM, the algebra of which induces their brackets to be those of Eqs (\ref{FINALDB}%
). These transformations also allow the canonical Hamiltonian density of the
NANM to be rewritten in terms of $A_{i}^{a}$ and  $E_{j}^{b}$, yielding a
result with exactly the same form as Eq. (\ref{hamiltonian_standar}).
Nevertheless, there is a main difference between the so-constructed
Hamiltonian and the full YM Hamiltonian. In fact, the Gauss laws $\Omega
^{a}=0$ do not appear as constraints in the NANM (since the $A_{0}^{a}$ are not
Lagrange multipliers), but rather as  functions of the respective d.o.f..

This has motivated us to search for a unified and simpler discussion of the
generic NANM. To this end, we find it convenient to generalize the
parametrization (\ref{NANM_LL}) to all cases in the form%
\begin{equation}
A_{0}^{a}=B^{a}\left( 1+\frac{N}{4B^{2}}\right)
,\,\,\,\,\,\,\,\,\,\,\,\,\,\,\,A_{3}^{a}=B^{a}\left( 1-\frac{N}{4B^{2}}%
\right) ,\;\;\;\;\;N=\left( A_{\bar{\imath}}^{b}A_{\bar{\imath}%
}^{b}+n^{2}M^{2}\right) ,\;\;\;4B^{2}\pm N\neq 0,\;\;\;  \label{GEN_PARAM}
\end{equation}%
which certainly satisfies the condition (\ref{NACOND}) and is written in
terms of the $3N$ independent GBs ($B^{a}$ and $A_{\bar{\imath}}^{b}$).

The relation between the above parametrization and the purely Goldstonic
d.o.f. introduced in Ref. \cite{JLCH2}, which we relabel as $\mathfrak{a}_{\mu
}^{b}$, can be established as follows. The $4N$ fields $\mathfrak{a}_{\mu
}^{b}$ are subjected to the $N$ additional constraints 
\begin{equation}
n^{\mu }\mathfrak{a}_{\mu }^{b}=0,  \label{ADDC-CH}
\end{equation}%
leaving only $3N$ independent GB modes. In terms of them, the original fields
are written as 
\begin{equation}
A_{\mu }^{b}=\mathfrak{a}_{\mu }^{b}-\frac{\mathfrak{n}_{\mu }^{b}}{%
\mathfrak{n}^{2}}\left( M^{2}-\mathfrak{n}^{2}\mathfrak{a}^{2}\right)
^{1/2},\;\;\mathfrak{n}_{\mu }^{b}=n_{\mu }s^{b},\;s^{2}=1,\;n^{2}\neq 0,\;
\label{PARAM_CH}
\end{equation}%
which satisfy the condition (\ref{NANMC}). Equation (\ref{PARAM_CH}) can be
inverted to produce%
\begin{equation}
\mathfrak{a}_{\mu }^{b}=A_{\mu }^{b}+\frac{\mathfrak{n}_{\mu }^{b}}{%
\mathfrak{n}^{2}}\left( \mathfrak{n\cdot }A\right) ,
\end{equation}%
which allows us to express $\mathfrak{a}_{\mu }^{b}=\mathfrak{a}_{\mu
}^{b}(B^{c},A_{\bar{\imath}}^{a})\;$by employing Eq.\ (\ref{GEN_PARAM}).

\bigskip

\subsection{The equations of motion}

After the substitution of (\ref{GEN_PARAM}) into the Lagrangian density (\ref%
{LNANM}), the variation of the corresponding action with respect to $A_{\nu
}^{a}$ yields 
\begin{equation}
0=\int d^{4}x(D_{\mu }F^{\mu \nu }-J^{\nu })^{a}\delta A_{\nu }^{a},
\label{ecumov}
\end{equation}%
where the $\delta A_{\nu }^{a}$ are not all independent. In our case, Eq. (%
\ref{GEN_PARAM}) leads to 
\begin{equation}
\delta A_{0}^{a}=\left[ \left( 1+\frac{N}{4B^{2}}\right) \delta ^{ab}-\frac{N%
}{4B^{2}}\frac{2B^{a}B^{b}}{B^{2}}\right] \delta B^{b}+\frac{B^{a}}{2B^{2}}%
A_{\bar{\imath}}^{b}\delta A_{\bar{\imath}}^{b},
\end{equation}
\begin{equation}
\delta A_{3}^{a}=\left[ \left( 1-\frac{N}{4B^{2}}\right) \delta ^{ab}+\frac{N%
}{4B^{2}}\frac{2B^{a}B^{b}}{B^{2}}\right] \delta B^{b}-\frac{B^{a}}{2B^{2}}%
A_{\bar{\imath}}^{b}\delta A_{\bar{\imath}}^{b},
\end{equation}%
in terms of the independent variations $\delta A_{\bar{\imath}}^{b}$ and $%
\delta B^{a}$. In this way the equations of motion are 
\begin{equation}
\delta A_{\bar{\imath}}^{a}:\;\;\;\;\mathcal{E}^{{\bar{\imath}}a}+\frac{B^{b}%
}{2B^{2}}\left[ \mathcal{E}^{0b}-\mathcal{E}^{3b}\right] A_{\bar{\imath}%
}^{a}=0,  \label{ecumov2}
\end{equation}
\begin{equation}
\delta B^{a}:\;\;0=\left( \left( 1+\frac{N}{4B^{2}}\right) \delta ^{ab}-%
\frac{N}{4B^{2}}\frac{2B^{b}B^{a}}{B^{2}}\right) \mathcal{E}^{0b}+\left(
\left( 1-\frac{N}{4B^{2}}\right) \delta ^{ab}+\frac{N}{4B^{2}}\frac{%
2B^{b}B^{a}}{B^{2}}\right) \mathcal{E}^{3b},  \label{ecumov3}
\end{equation}%
in the notation of Eq. (\ref{EQMOTYM}).

Let us recall that in the case of the SO$(N)$ YM theory the equations of
motion are just given by $\mathcal{E}^{\nu a}=0$. Also, the above equations
of motion do not imply current conservation $D_{\nu }J^{\nu a}=0$, basically
because the condition (\ref{NACOND}) breaks non-Abelian gauge invariance. A
way to recover the YM equations of motion together with gauge invariance is
to impose the Gauss laws $\mathcal{E}^{0a}=0$. In this way, under the
conditions $4B^2\pm N \neq 0$, Eq. (\ref{ecumov3}) yields the solution $%
\mathcal{E}^{3b}=0$. These two conditions in Eq. (\ref{ecumov2}) provide the
final set $\mathcal{E}^{{\bar{\imath}}a}=0.$

\subsection{The Hamiltonian density}

In order to unify the notation when going to the Hamiltonian formulation we
introduce the $3N$ d.o.f. $\Phi _{A}^{a},\;A=1,2,3,$ 
\begin{equation}
\Phi _{1}^{a}=A_{1}^{a},\;\;\;\;\Phi _{2}^{a}=A_{2}^{a}\;,\;\;\;\Phi
_{3}^{a}=B^{a},\;
\end{equation}%
in such a way that the coordinate transformation 
\begin{equation}
A_{i}^{a}=A_{i}^{a}(\Phi _{A}^{b}),  \label{NACHVAR}
\end{equation}%
arising from Eq. (\ref{GEN_PARAM}) is invertible. In fact, the inverses are 
\begin{equation}
\Phi _{1}^{a}=A_{1}^{a},\;\;\;\;\Phi _{2}^{a}=A_{2}^{a},\;\;\;\;\;\;\Phi
_{3}^{a}=\frac{A_{3}^{a}}{2\sqrt{A_{3}^{b}A_{3}^{b}}}\left( \sqrt{%
A_{3}^{b}A_{3}^{b}}+\sqrt{A_{i}^{b}A_{i}^{b}+n^{2}M^{2}}\right) .
\label{NAINVERSE}
\end{equation}%
We also have 
\begin{equation}
A_{0}^{a}=A_{0}^{a}(\Phi _{A}^{b}),  \label{A0PHIA}
\end{equation}%
in terms of Eq. (\ref{NACHVAR}), according to the first relation in Eq. (\ref%
{GEN_PARAM}). The relevant property of the transformation (\ref{NACHVAR}) is
that 
\begin{equation}
\dot{A}_{i}^{a}=\frac{\partial A_{i}^{a}}{\partial \Phi _{B}^{b}}\dot{\Phi}%
_{B}^{b} \;\;\;\;\rightarrow \;\;\;\frac{\partial \dot{A}_{i}^{a}}{\partial 
\dot{\Phi}_{B}^{b}}=\frac{\partial A_{i}^{a}}{\partial \Phi _{B}^{b}},
\label{FUNDCT}
\end{equation}%
together with the invertibility of the velocities 
\begin{equation}
\dot{\Phi}_{A}^{a}=\frac{\partial \Phi _{A}^{a}}{\partial A_{i}^{b}}\dot{A}%
_{i}^{b}.  \label{INVERTVEL}
\end{equation}

In the following we will not require the explicit form of the
transformations (\ref{GEN_PARAM}) and (\ref{NAINVERSE}), but only their
generic form (\ref{NACHVAR}), together with the property that this
transformation can be inverted.

Next we proceed to calculate the Hamiltonian density of the NANM in terms of
the canonically conjugated variables $\Phi _{A}^{b}$ and $\Pi _{A}^{b}$, and
we employ a procedure that allows us to make direct contact with both the YM
Hamiltonian density (\ref{hamiltonian_standar}) and the YM canonical algebra
(\ref{FINALDB}). After making the substitutions (\ref{NACHVAR}) and (\ref%
{A0PHIA}), the Lagrangian density (\ref{LNANM}) can be rewritten as 
\begin{equation}
\mathcal{L}_{\rm NANM}(\Phi ,\dot{\Phi})=\frac{1}{2}E_{i}^{a}E_{i}^{a}-\frac{1}{2%
}B_{i}^{a}B_{i}^{a}-J^{a\mu }A_{\mu }^{a},\,
\end{equation}%
where 
\begin{equation}
E_{i}^{a}=\dot{A}_{i}^{a}-D_{i}A_{0}^{a},\;\;\;\;\;\;\;B_{i}^{a}=\frac{1}{2}%
\epsilon _{ijk}F_{jk}^{a},\;  \label{COLOREB}
\end{equation}%
with $E_{i}^{a}=E_{i}^{a}(\Phi ,\dot{\Phi})$ and $B_{i}^{a}=B_{i}^{a}(\Phi )$%
. The canonically conjugate momenta are calculated as 
\[
\Pi _{A}^{a}=\frac{\partial \mathcal{L}_{\rm NANM}(\Phi ,\dot{\Phi})}{\partial 
\dot{\Phi}_{A}^{a}}=E_{i}^{b}\frac{\partial \dot{A}_{i}^{b}}{\partial \dot{%
\Phi}_{A}^{a}}=E_{i}^{b}\frac{\partial A_{i}^{b}}{\partial \Phi _{A}^{a}}, 
\]%
employing (\ref{FUNDCT}). The inverse of Eq. (\ref{NACHVAR}) allows us
to write the colored electric fields $E_{i}^{a}$ as functions of the momenta 
$\Pi _{A}^{b}$ of the NANM%
\begin{equation}
E_{i}^{b}(\Phi ,\Pi )=\frac{\partial \Phi _{A}^{a}}{\partial A_{i}^{b}}\Pi
_{A}^{a}.  \label{MOMCT}
\end{equation}%
The Wronskian of the system is 
\begin{equation}
\det \left( \frac{\partial ^{2}\mathcal{L}_{\rm NANM}(\Phi ,\dot{\Phi})}{%
\partial \dot{\Phi}_{A}^{a}\partial \dot{\Phi}_{B}^{b}}\right) =\det \left( 
\frac{\partial \Pi _{A}^{a}}{\partial \dot{\Phi}_{B}^{b}}\right) =\det
\left( \frac{\partial \dot{A}_{i}^{c}}{\partial \dot{\Phi}_{B}^{b}}\frac{%
\partial A_{i}^{c}}{\partial \Phi _{A}^{a}}\right) =\det \left( \frac{%
\partial A_{i}^{c}}{\partial \Phi _{B}^{b}}\frac{\partial A_{i}^{c}}{%
\partial \Phi _{A}^{a}}\right) \neq 0, 
\end{equation}
as shown in Appendix C. In this way, the NANM is exhibited as a regular
system in the parametrization (\ref{GEN_PARAM}), so that no constraints are
present.

The NANM Hamiltonian density is 
\begin{equation}
\mathcal{H}_{\rm NANM}=\;\Pi _{A}^{a}\dot{\Phi}_{A}^{a}-\left( \frac{1}{2}%
E_{i}^{a}E_{i}^{a}-\frac{1}{2}B_{i}^{a}B_{i}^{a}-J^{a\mu }A_{\mu
}^{a}\right) ,
\end{equation}%
which we rewrite in successive steps 
\begin{eqnarray}
\mathcal{H}_{\rm NANM} &=&\;\Pi _{A}^{a}\frac{\partial \Phi _{A}^{a}}{\partial
A_{i}^{b}}\dot{A}_{i}^{b}-\left( \frac{1}{2}E_{i}^{a}E_{i}^{a}-\frac{1}{2}%
B_{i}^{a}B_{i}^{a}-J^{a\mu }A_{\mu }^{a}\right) , \\
\mathcal{H}_{\rm NANM} &=&\;E_{i}^{b}\dot{A}_{i}^{b}-\left( \frac{1}{2}%
E_{i}^{a}E_{i}^{a}-\frac{1}{2}B_{i}^{a}B_{i}^{a}-J^{a\mu }A_{\mu
}^{a}\right),  \label{HNANM_INT} \\
\mathcal{H}_{\rm NANM} &=&\;E_{i}^{b}\left( E_{i}^{b}+D_{i}A_{0}^{b}\right)
-\left( \frac{1}{2}E_{i}^{a}E_{i}^{a}-\frac{1}{2}B_{i}^{a}B_{i}^{a}-J^{a\mu
}A_{\mu }^{a}\right), \\
\mathcal{H}_{\rm NANM}(\Phi ,\Pi ) &=&\;\frac{1}{2}E_{i}^{a}E_{i}^{a}+\frac{1}{2}%
B_{i}^{a}B_{i}^{a}-\left( D_{i}E_{i}^{b}-J^{b0}\right)
A_{0}^{b}+J^{ai}A_{i}^{a},  \label{HNANM_FIN}
\end{eqnarray}%
where we have used Eqs. (\ref{INVERTVEL}), (\ref{COLOREB}), and (\ref%
{MOMCT}), together with an integration by parts in the term containing the
covariant derivative. The dependence of $\mathcal{H}_{\rm NANM}$ on the
canonical variables $\Phi ,\Pi$ is clearly established by the change of
variables (\ref{NACHVAR}), (\ref{A0PHIA}) \ and (\ref{MOMCT}). The canonical variables of NANM
satisfy the standard PBs %
\begin{equation}
\left\{ \Phi _{A}^{a}(\mathbf{x}),\Phi _{B}^{b}(\mathbf{y})\right\}
=0,\;\;\;\left\{ \Pi _{A}^{a}(\mathbf{x}),\Pi _{B}^{b}(\mathbf{y})\right\}
=0,\;\;\;\;\left\{ \Phi _{A}^{a}(\mathbf{x}),\Pi _{B}^{b}(\mathbf{y}%
)\right\} =\delta ^{ab}\delta _{AB}\delta ^{3}(\mathbf{x-y}).  \label{NANMPB}
\end{equation}%
Now we can consider the NANM Hamiltonian density\ (\ref{HNANM_FIN}) from the
perspective of the fields $A_{i}^{a}$ and $E_{i}^{a}$. The relation
arising from the velocity-dependent term of the NANM\ Hamiltonian action, 
\begin{equation}
\int d^{4}x\;\Pi _{A}^{a}\dot{\Phi}_{A}^{a}=\int d^{4}x\;E_{i}^{a}\dot{A}%
_{i}^{a}= \int d^{4}x\;(-E^{ai})\dot{A}_{i}^{a} ,
\end{equation}
[used previously in obtaining Eq. (\ref{HNANM_INT})], establishes $(-E^{ai})$
as the canonically conjugate momenta of $A_{i}^{a}$. In this way Eq. (\ref%
{HNANM_FIN}) can be seen as a Hamiltonian density $\mathcal{H}(A,E)\;$%
obtained from $\mathcal{H}_{\rm NANM}(\Phi ,\Pi )$ via the substitution of the
phase-space transformations 
\begin{equation}
(\Phi ,\Pi )\rightarrow (A,E),  \label{PSTRANS}
\end{equation}%
which follow from the inverses of Eqs. (\ref{NACHVAR}) and (\ref{MOMCT}) plus
Eq. (\ref{A0PHIA}), 
\begin{equation}
A_{0}^{a}=\frac{A_{3}^{a}}{\sqrt{A_{3}^{b}A_{3}^{b}}}\left( \sqrt{%
A_{i}^{b}A_{i}^{b}+n^{2}M^{2}}\right) ,  \label{A0INV}
\end{equation}%
in terms of the new variables. But, since the transformations (\ref{MOMCT})
are generated by the change of variables (\ref{NACHVAR}) in coordinate
space, we know from classical mechanics that the full transformation in
phase space is a canonical transformation. In this way we automatically
recover the PBs
\begin{equation}
\left\{ A_{i}^{a}(\mathbf{x}),A_{j}^{b}(\mathbf{y})\right\} =0,\;\;\;\left\{
E^{ai}(\mathbf{x}),E^{bj}(\mathbf{y})\right\}=0,\;\;\;\;\left\{ A_{i}^{a}(%
\mathbf{x}),E^{bj}(\mathbf{y})\right\} =-\delta ^{ab}\delta
_{i}^{j}\delta^{3}(\mathbf{x-y})  \label{AEPBALG}
\end{equation}%
from Eq. (\ref{NANMPB}). To summarize, from each Hamiltonian version of the
NANM, defined by the different values of $n^{2}$, we can regain (via a
canonical transformation) the Hamiltonian density (\ref{HNANM_FIN}) together
with the canonical algebra (\ref{FINALDB}). The Hamiltonian density (\ref%
{HNANM_FIN}) differs from the YM Hamiltonian density (\ref%
{hamiltonian_standar}) only in the fact that the Gauss laws $\Omega
^{b}=\left( D_{i}E_{i}^{b}-J^{b0}\right) =0$ do not appear as first-class
constraints, because $A_{0}^{a}$ are not arbitrary Lagrange multipliers, but
rather as functions of the coordinates, as shown in Eq. (\ref{A0INV}).

\subsection{The evolution of the Gauss functions\ $\Omega ^{a}$}

The calculation follows the same steps dictated by Eq. (\ref{TEVOLGF}) in
the case of the SL-NANM. In the parametrization (\ref{GEN_PARAM}), the
result is 
\begin{equation}
\dot{\Omega}^{a}=-gC^{abc}A_{0}^{b}\Omega ^{c}-D_{\mu }J^{\mu a}+D_{3}\left( 
\frac{A_{0}^{1}}{A_{3}^{1}}\Omega ^{a}\right) -D_{3}\left( \frac{A_{3}^{a}}{%
A_{3}^{c}A_{3}^{c}}A_{0}^{b}\Omega ^{b}\right) +D_{i}\left( \frac{A_{i}^{a}}{%
N}A_{0}^{b}\Omega ^{b}\right),
\end{equation}%
where $A_{0}^{b}\;$is given by Eq. (\ref{A0INV}). The above evolution
equation leads to the same statements regarding the equivalence of the 
SO$(N)$ YM theory with the NANM as those stated after Eq. (\ref%
{RESULTSLNANM}) in the case of the SL-NANM.

\section{Summary and Final Comments}

The possible interpretation of gauge particles (e.g., photons and gravitons) as the GB modes arising from some
spontaneous symmetry breaking is an interesting hypothesis that would
provide a dynamical setting for the gauge principle.

In this paper we have taken the Nambu approach, whereby spontaneous
SLSB is incorporated in an effective way in the
model by means of a nonlinear constraint. These models can be understood
as generalizations of the nonlinear sigma model describing pion
interactions. The challenge posed by this setting is to show the
conditions under which a the violations of Lorentz symmetry and of gauge invariance
(introduced by the nonlinear constraint) are  unobservable in such a way that
the Goldstone bosons that appear can be interpreted as the gauge particles of
an unbroken gauge theory. In other words, one tries to determine the
conditions under which the corresponding Nambu model is equivalent to the
unbroken gauge theory. Such conditions have been studied using perturbation
theory for electrodynamics and YM theories, for example, in Refs.
\cite{Nambu-Progr, Azatov-Chkareuli, JLCH1,JLCH2, JLCH3}. The main result is
that, to the order considered (usually the tree-level or one-loop corrections)
and after imposing the Gauss laws plus current conservation, the violations
of Lorentz symmetry are unobservable, so that the corresponding
Nambu model reproduces the corresponding  gauge theory, with the gauge bosons
realized as the corresponding Goldstone bosons.

In this work we have generalized the nonperturbative Hamiltonian
analysis developed for the Abelian Nambu model in Ref.  \cite{Urru-Mont} to the non-Abelian case.
Also, we have made an important conceptual and practical improvement in the
method  of dealing with the relation between the NANM and the corresponding YM
theory. On the other hand, no discussion is provided here about the
possible perturbative equivalence of the corrected Nambu model and the YM
theory in a fixed gauge.

In Sec. II we considered the specific case of the SL-NANM by solving the nonlinear constraint (\ref{NANMC}) in
terms of $A_{\mu =3}^{a=1}$ and starting with the remaining $4N-1$ d.o.f. per
point in coordinate space. The Lagrangian equations of motion were obtained,
yielding different results than from the YM equations of motion, as
expected. The canonical momenta and the canonical Hamiltonian were subsequently
constructed, with the appearance of $2(N-1)$ second-class constraints. The
standard variables of the YM theory, $A_{i}^{a}$ and $E^{bj},i=1,2,3$, were
written in terms of the canonical variables of the SL-NANM, the canonical
algebra of which  induces the standard YM algebra for the former variables at the
level of PBs. Appendix B includes a summary of the
required PBs which  prove the previous statement. The second-class constraints
were further strongly imposed  by introducing Dirac brackets, whose values
for the variables $A_{i}^{a}$ and $E^{bj}$ turned out to be the same as the
previously calculated PBs. The final extended Hamiltonian for the SL-NANM,
rewritten in terms of the variables $A_{i}^{a}$ and $E^{bj}$, has the same
form as the standard YM Hamiltonian, except that the Gauss laws $\Omega
^{a}=0$ do not appear as first-class constraints. The time evolution of the
functions $\Omega ^{a}$, according to the SL-NANM dynamics, were calculated,
yielding the result that after demanding current conservation, the
imposition of the Gauss laws at some initial time yields $\Omega ^{a}=0\;$%
for all times. The final $3N$ d.o.f. in coordinate space of the NANM were
recovered, since $\frac{1}{2}\left( 2(4N-1)-2(N-1)\right) =3N$. It was
emphasized that a similar analysis could be carried out for each of the
remaining cases corresponding to the time-like and light-like versions of
the NANM.

Section III presented a substantial conceptual and practical improvement over
the previous individual calculations to study the relation between NANM
and YM theories. Starting from an alternative parametrization that solves
the constraint (\ref{NANMC}) for arbitrary values of $n^{2}$ in terms of $3N
$ d.o.f. $\Phi _{A}^{a},\; A=1,2,3$, we showed that the calculation of the NANM
canonical momenta $\Pi _{A}^{a}$ can be written in such a way that the
chosen parametrization induces a direct relation between the YM variables $%
A_{i}^{a}$ and $E^{bj}$ and the canonical variables of the NANM. Since the
phase-space transformations are induced by a coordinate transformation (the
chosen parametrization), we know from classical mechanics that the phase-space transformation $(\Phi _{A}^{a},\Pi _{B}^{b})\rightarrow
(A_{i}^{a},E^{bj})$ is a canonical transformation, once the $E^{bj}$ are
recognized as the momenta canonically conjugate to the $A_{i}^{a}$ via the
kinetic part of the NANM Hamiltonian action. In this way, one immediately
concludes that the resulting algebra of the $(A_{i}^{a},E^{bj})$ has to
be the canonical one, without requiring the detailed and tedious
calculations that were necessary in the discussion of the previous section.
It is interesting to observe that the identification of the canonical
transformation is independent of the detailed structure of the chosen
parametrization, as soon as it provides an invertible change of coordinates 
$\Phi _{A}^{a}=\Phi _{A}^{a}(A_{i}^{a})$. Another useful property of this
parametrization is that it exhibits the NANM as a regular theory (i.e., no
constraints appear in the Hamiltonian analysis), since it includes just the
necessary $3N$ d.o.f. of the NANM. This was proved in Appendix C. The
canonical NANM Hamiltonian, rewritten in terms of the YM variables $%
(A_{i}^{a},E^{bj})$, again has the same form as the YM Hamiltonian, except
that the Gauss laws do not arise as first-class constraints. The time
evolution of the functions $\Omega ^{a}$ was also calculated, with similar
results as in the SL-NANM.

The relation between our approach and the method of Ref. \cite{JLCH2},\
which also included pure Goldstone field modes, was elucidated in the
paragraphs after Eqs. (\ref{HAMFIN_SLNANM}) and (\ref{GEN_PARAM}).

To summarize, a nonperturbative equivalence between the SO$(N)$ YM
 theory and the corresponding NANM has been established, after current
conservation and the Gauss laws are imposed as initial conditions for the
latter. Actually, the Gauss laws $-$ valid now for all times $-$ are next
added as  Hamiltonian constraints $-N^{a}\Omega ^{a}$ to Eq. (\ref%
{HNANM_FIN}), with arbitrary functions $N^{a}$. The further redefinition  $%
A_{0}^{a}+N^{a}=\Theta ^{a}$  leads to the final YM Hamiltonian%
\begin{equation}
\mathcal{H}_{\rm YM}=\frac{1}{2}(\mathbf{E}^{2}+\mathbf{B}^{2})-\Theta ^{a}\Omega
^{a}+J_{i}^{a}A^{ia},
\end{equation}%
where $\Theta ^{a}$ are now arbitrary functions. In other words, the Gauss
laws are imposed {\it \`{a} la} Dirac upon the physical states $|\Psi \rangle _{\rm phys}
$ by demanding that   $\Omega ^{a}|\Psi \rangle _{\rm phys}=0$. Also, the
emergence of the SO$(N)$ YM theory subsequently guarantees current
conservation for all times, as a consequence of the YM equations of motion. The
established  equivalence is independent of any gauge fixing and supports the
idea that gauge particles arise as the Goldstone bosons of a model
exhibiting a spontaneous Lorentz symmetry breaking that is not physically
observable.

\section*{Acknowledgements}

L. F. U. was partially supported by projects UNAM (Direcci\'on General de Aduntos del Personal Acad\'emico) \#  IN104815 and CONACyT \#
237503. He acknowledges very useful discussions with J. D. Vergara and J. A.
Garc\'\i a-Zenteno.

\appendix

\section{The SO(N) Yang--Mills theory}

\label{APPA}

We present a brief review of the Hamiltonian formulation of the standard 
SO$(N)$ YM theory. The main motivation, besides establishing some
notation, is to recall the basic properties of the YM theory that have to be
recovered in order to state its emergence from the different versions of the
NANM.

The Yang-Mills Lagrangian density is given by 
\begin{equation}
\mathcal{L}=Tr\left[ -\frac{1}{4}\mathbf{F}_{\mu \nu }\mathbf{F}^{\mu \nu }-%
\mathbf{J}^{\mu }\mathbf{A}_{\mu }\right] ,  \label{LDENS}
\end{equation}%
where boldfaced quantities denote matrices in the Lie algebra of the
internal symmetry group SO$(N)$ with $N(N-1)/2$ generators $t^{a}$; i.e., $%
\mathbf{M}=M^{a}t^{a}$. This algebra is generated by $\;\left[ t^{a},t^{b}%
\right] =C^{abc}t^{c}$, where the structure constants $C^{abc}$ are
completely antisymmetric. The field strength is 
\begin{equation}
\mathbf{F}_{\mu \nu }=\partial _{\mu }\mathbf{A}_{\nu }-\partial _{\nu }%
\mathbf{A}_{\mu }+g\left[ \mathbf{A}_{\mu },\mathbf{A}_{\nu }\right] ,
\label{FS}
\end{equation}%
and the equations of motion are 
\begin{equation}
D_{\mu }\mathbf{F}^{\mu \nu }=\mathbf{J}^{\nu },  \label{EQMOT}
\end{equation}%
where the covariant derivative is defined as 
\begin{equation}
D_{\mu }\mathbf{M}=\partial _{\mu }\mathbf{M\;+}g\left[ \mathbf{A}_{\mu },%
\mathbf{M}\right] .  \label{COVDER}
\end{equation}%
From the above definitions we obtain%
\begin{equation}
\left[ D_{\mu },D_{\nu }\right] \mathbf{M}=\left[ \mathbf{M,\mathbf{F}_{\mu
\nu }\;}\right] ,  \label{REL1}
\end{equation}%
which leads to current conservation, $D_{\nu }\mathbf{J}^{\nu }=0.$

The expressions of Eqs. (\ref{FS}) and (\ref{COVDER}) in terms of the
components of the corresponding fields are 
\begin{eqnarray}
F_{\mu \nu }^{a} &=&\partial _{\mu }A_{\nu }^{a}-\partial _{\nu }A_{\mu
}^{a}+gC^{abc}A_{\mu }^{b}A_{\nu }^{c},  \label{tensor} \\
\left( D_{\mu }M\right) ^{a} &\equiv &D_{\mu }M^{a}=\partial _{\mu
}M^{a}+gC^{abc}A_{\mu }^{b}M^{c}.
\end{eqnarray}%
The Jacobi identity for the connection $\mathbf{A}_{\mu }\;$is%
\begin{equation}
C^{arb}C^{cdr}+C^{crb}C^{dar}+C^{drb}C^{acr}=0  \label{JI}
\end{equation}%
in terms of the structure constants. The group indices $a=1,2,...,N\,$\ are
raised (lowered) by the metric $\delta ^{ab}\;(\delta _{ab})$\ and their
position as superscripts or subscripts is just a matter of convenience in
writing the corresponding expression.\ 

Next we review the Hamiltonian version of the YM theory. The canonical
momenta are given by 
\begin{equation}
\Pi ^{a\mu }=\frac{\partial L}{\partial (\dot{A}_{\mu }^{a})}.
\end{equation}%
Therefore, considering 
\begin{equation}
\frac{\partial (F_{\mu \nu }^{a}F_{a}^{\mu \nu })}{\partial (\dot{A}_{\alpha
}^{a})}=4F_{a}^{0\alpha },  \nonumber
\end{equation}%
we find 
\begin{equation}
\Pi _{0}^{a}=0,\,\,\,\,\,\,\,\,\,\,\,\,\,\,\;\;\;\;\,\,\,\,\Pi
_{i}^{a}=F_{i0}^{a}\equiv -E_{i}^{a},
\end{equation}%
which satisfy the nonzero PBs 
\begin{equation}
\;\;\{A_{0}^{a}(\mathbf{x},t),\Pi _{0}^{b}(\mathbf{y},t)\}=\delta
^{ab}\delta (\mathbf{x}-\mathbf{y}),\;\;\;\;\;\{A_{i}^{a}(\mathbf{x},t),\Pi
^{bj}(\mathbf{y},t)\}=\delta _{i}^{j}\delta ^{ab}\delta (\mathbf{x}-\mathbf{y%
}).  \label{PB1}
\end{equation}%
In the following we assume that all PBs are calculated at equal times and we
suppress the label $t$ in most cases. From Eq. (\ref{tensor}) we get $\dot{A}%
_{i}^{a}$ as 
\begin{equation}
\dot{A}_{i}^{a}=E_{i}^{a}+\partial
_{i}A_{0}^{a}+gC^{abc}\,A_{i}^{b}A_{0}^{c}=E_{i}^{a}+D_{i}A_{0}^{a}.
\label{CONV_NOTE}
\end{equation}%
We also introduce 
\begin{equation}
B_{k}^{a}=\frac{1}{2}\epsilon ^{ij}\,_{k}F_{ij}^{a}.
\end{equation}
Recalling that 
\begin{equation}
D_{\mu }(N^{a}M^{a})=\left( D_{\mu }N^{a}\right) M^{a}+N^{a}(D_{\mu
}M^{a})=\partial _{\mu }(N^{a}M^{a}),\;\;\;a=1,2,...,N,
\end{equation}%
which allows us to perform integration by parts within the action, we find the
canonical Hamiltonian density 
\begin{equation}
\mathcal{H}=\Pi _{a}^{i}\dot{A}_{i}^{a}-\mathcal{L}=\frac{1}{2}(\mathbf{E}%
^{2}+\mathbf{B}^{2})-A_{0}^{a}(D_{i}E_{i}-J^{0})^{a}-J_{i}^{a}A_{i}^{a},
\end{equation}%
where%
\begin{equation}
\mathbf{E}^{2}=tr(\mathbf{F}_{0i}\mathbf{F}_{0i}),\qquad \mathbf{B}^{2}=%
\frac{1}{2}tr(\mathbf{F}_{ij}\mathbf{F}^{ij}).
\end{equation}

We employ Dirac's method to construct the canonical theory, due to the
fact that primary constraints 
\begin{equation}
\Sigma ^{a}=\Pi _{0}^{a}\simeq 0,  \label{YMC1}
\end{equation}%
are present. The extended Hamiltonian density is given by 
\begin{equation}
\mathcal{H}_{E}=\frac{1}{2}(\mathbf{E}^{2}+\mathbf{B}%
^{2})-A_{0}^{a}(D_{i}E_{i}-J^{0})_{a}+J_{i}^{a}A_{a}^{i}+\lambda ^{a}\Sigma
_{a},
\end{equation}%
where $\lambda ^{a}$ are arbitrary functions. The evolution condition of the
primary constraints 
\begin{equation}
\dot{\Sigma}^{a}(\mathbf{x})=\{\Sigma ^{a}(\mathbf{x}),\int d^{3}y\;H_{E}(%
\mathbf{y})\}\simeq 0,
\end{equation}%
leads to the Gauss laws 
\begin{equation}
\Omega ^{a}=(D_{i}E_{i}-J^{0})^{a}\simeq 0.  \label{YMC2}
\end{equation}%
It is not difficult to prove that Eqs. (\ref{YMC1}) \ and (\ref{YMC2}) are the
only constraints present and that they constitute a first-class set. In
fact, calculating the time evolution of $\Omega ^{a}$ yields%
\begin{equation}
\dot{\Omega}^{a}(\mathbf{x})=\left\{ \Omega ^{a}(\mathbf{x}),\;\int d^{3}y\;%
\mathcal{H}_{E}(\mathbf{y})\right\} =\left\{ \Omega ^{a}(\mathbf{x}),\;\int
d^{3}y\;\left( \frac{1}{2}\mathbf{B}^{2}-A_{0}^{b}\Omega
^{b}+J_{i}^{a}A_{a}^{i}\right) \right\} -\partial _{0}J_{0}^{a}.
\end{equation}
From the PBs (\ref{PB1}) together with Eq. (\ref{JI})\ we obtain%
\begin{equation}
\int d^{3}y\left\{ \Omega ^{a}(\mathbf{x}),\mathbf{B}^{2}(\mathbf{y}%
)\right\} =0.  \label{OMEGABB}
\end{equation}%
The PB of the constraints (\ref{YMC2}) produces 
\begin{equation}
\left\{ D_{i}E_{i}^{a}(\mathbf{x}),\;\int d^3y\;D_{j}E_{j}^{b}(\mathbf{y}%
)\right\} = C^{abc}D_{k}E_{k}^{c}(\mathbf{x}),
\end{equation}%
which leads to 
\begin{equation}
\int d^{3}y\;\left\{ \Omega ^{a}(\mathbf{x}),\Omega ^{b}(\mathbf{y})\right\}
M^{b}(\mathbf{y})=C^{abc}M^{b}\left[ \Omega ^{c}+J_{0}^{c}\right] (\mathbf{x}%
).  \label{PPGAUSS1}
\end{equation}%
In this way, 
\begin{equation}
\dot{\Omega}^{a}=-\partial
_{0}J_{0}^{a}-C^{abc}A_{0}^{b}J_{0}^{c}-C^{abc}A_{0}^{b}\Omega
^{c}-D_{k}J^{ak}=-C^{abc}A_{0}^{b}\Omega ^{c}-D_{\mu }J^{a\mu },
\end{equation}
which is zero, modulo the constraints and using current conservation.

Normally one fixes 
\begin{equation}
\Pi _{0}^{a}\simeq 0,\,\,\,\,\,\,\,\,\,A_{0}^{a}\simeq \Theta^a ,
\label{cond_YM}
\end{equation}%
with $\Theta^{a}$ being arbitrary functions to be consistently determined
after the remaining first-class constraints $\Omega ^{a}\;$are fixed. $\;$

The final Hamiltonian density is 
\begin{equation}
\mathcal{H}_{E}=\frac{1}{2}(\mathbf{E}^{2}+\mathbf{B}^{2})-\Theta
^{a}(D_{i}E_{i}-J^{0})_{a}+J_{i}^{a}A_{a}^{i}.  \label{hamiltonian_standar}
\end{equation}%
Once $\Pi _{0}^{a}$ and $\,A_{0}^{a}$ are strongly fixed, the Dirac
brackets of the remaining variables are 
\begin{equation}
\{A_{i}^{a}(\mathbf{x},t),A_{j}^{b}(\mathbf{y},t)\}^{\ast
}=0,\,\,\,\,\,\,\,\,\,\,\,\,\{E^{ai}(\mathbf{x},t),E^{bj}(\mathbf{y}%
,t)\}^{\ast }=0,\;\;\;\;\{A_{i}^{a}(\mathbf{x},t),E^{bj}(\mathbf{y}%
,t)\}^{\ast }=-\delta _{i}^{j}\delta ^{ab}\delta (\mathbf{x}-\mathbf{y}).
\label{FINALDB}
\end{equation}
The final count of the number of d.o.f. per point in coordinate space
yields%
\begin{equation}
\#d.o.f.=\frac{1}{2}(2\times 4N-2\times 2N)=2N.
\end{equation}

\section{ The bracket algebra in the SL-NANM}

In this case the canonical variables are $\;A_{0}^{a},\;A_{\bar{\imath}%
}^{1},\;A_{i}^{\bar{a}},\;\Pi _{0}^{a},\;\Pi ^{1\bar{k}},\;\Pi ^{\bar{a}i}\;$%
where $a=1,2,...,N;\;\bar{a}=2,3,...,N;\;i=1,2,3\;$and $\bar{\imath}=1,2.\;$%
The nonzero PBs are%
\begin{eqnarray}
\{A_{0}^{a}(\mathbf{x},t),\Pi _{0}^{b}(\mathbf{y},t)\} &=&\delta ^{3}(%
\mathbf{x}-\mathbf{y})\delta ^{ab}\,\,\,,\;\;\;\;\;\;\;\{A_{\bar{\imath}%
}^{1}(\mathbf{x},t),\Pi ^{1\bar{k}}(\mathbf{y},t)\}=\delta _{\bar{\imath}}^{%
\bar{k}}\delta ^{3}(\mathbf{x}-\mathbf{y}).\,\,\,\,\,\,\,\,\,  \nonumber \\
\,\,\{A_{i}^{\bar{a}}(\mathbf{x},t),\Pi ^{\bar{b}k}(\mathbf{y},t)\}
&=&\delta _{i}^{k}\delta ^{\bar{a}\bar{b}}\delta ^{3}(\mathbf{x}-\mathbf{y}).
\label{SL_CA1}
\end{eqnarray}%
The YM canonical variables $(A_{i}^{a},E_{j}^{b})$ are given by%
\begin{eqnarray}
&& A_{3}^{1}=\sqrt{M^{2}+\left( A_{0}^{a}\right) ^{2}-\left( A_{\bar{\imath}%
}^{1}\right) ^{2}-\left( A_{k}^{\bar{a}}\right) ^{2}},\;\;\;A_{\bar{\imath}%
}^{1},\;A_{i}^{\bar{a}},  \nonumber \\
&& E_{3}^{1}=A_{3}^{1}\left( \frac{\Pi _{0}^{1}}{A_{0}^{1}}\right)
,\,\,\;\;E_{\bar{\imath}}^{1}=\Pi ^{1\bar{\imath}}+\left( \frac{\Pi _{0}^{1}%
}{A_{0}^{1}}\right) A_{\bar{\imath}}^{1}\;\,\,\,\,,\,E_{i}^{\bar{a}}=\Pi ^{%
\bar{a}i}+\left( \frac{\Pi _{0}^{1}}{A_{0}^{1}}\right) A_{i}^{\bar{a}},
\label{YMCVSL}
\end{eqnarray}%
in terms of the canonical variables of the SL-NANM.

In the following, we do
not specify the coordinate dependence of each term: it is to be understood
according to the following convention%
\begin{equation}
\left\{ P(\mathbf{x}),Q(\mathbf{y})\right\} =\left\{ P,Q\right\} .
\end{equation}%
Additionally, we also suppress the unit $\delta ^{3}(\mathbf{x-y})\;$%
in coordinate space. We do not provide any details for each derivation: we
only include the final results.

Our first goal is to calculate the equal-times algebra among the YM
variables $A_{i}^{a},\;E_{j}^{b}$ defined in Eq. (\ref{YMCVSL}) in terms of
the canonical algebra (\ref{SL_CA1}) of the SL-NANM. To this end we first consider  the PBs between the $A_{3}^{1}$ [which is the
solution of the constraint (\ref{NACOND})]  and the canonical conjugate
momenta of the SL-NANM. The results are 
\begin{equation}
\;\{\Pi ^{\bar{a}k},A_{3}^{1}\}=\frac{A_{k}^{\bar{a}}}{A_{3}^{1}}%
,\;\;\;\;\;\{\Pi _{0}^{a},A_{3}^{1}\}=-\frac{A_{0}^{a}}{A_{3}^{1}}%
,\;\;\;\;\;\;\{\Pi ^{1\bar{k}},A_{3}^{1}\}=\frac{A_{\bar{k}}^{1}}{A_{3}^{1}}.
\label{PIACOM}
\end{equation}

\subsection{The $A-A$ sector}

The algebra among the canonical coordinates $\left( A_{\bar{\imath}%
}^{1},\;A_{k}^{\bar{a}}\right)$ and $A_{3}^{1}$ is trivial because the
former satisfy the canonical relations  and $A_{3}^{1}=A_{3}^{1}(A_{\bar{\imath}%
}^{1},\;A_{k}^{\bar{a}})$, so that we have $\left\{
A_{i}^{a},A_{j}^{b}\right\} =0.$
\subsection{The $A-E$ sector}
We obtain
\begin{equation}
\left\{ A_{3}^{1},E_{3}^{1}\right\} =1,\;\;\;\left\{ A_{3}^{1},E_{\bar{k}%
}^{1}\right\} =0,\;\;\;\left\{ A_{3}^{1},E_{i}^{\bar{a}}\right\}
=0,\;\;\left\{ A_{\bar{\imath}}^{1},E_{3}^{1}\right\}=0,
\end{equation}%
\begin{equation}
\left\{ A_{\bar{\imath}}^{1},E_{\bar{k}}^{1}\right\} =\delta _{\bar{\imath}%
}^{\bar{k}},\;\;\;\left\{ A_{\bar{\imath}}^{1},E_{k}^{\bar{a}}\right\}
=0,\;\;\;\left\{ A_{i}^{\bar{a}},E_{3}^{1}\right\} =0,\;\;\;\left\{ A_{i}^{%
\bar{a}},E_{\bar{k}}^{1}\right\} =0,\;\left\{ A_{i}^{\bar{b}},E_{j}^{\bar{a}%
}\right\} =\delta ^{\bar{a}\bar{b}}\delta _{i}^{j}.\;
\end{equation}%
The above relations can be summarized as 
\begin{equation}
\left\{ A_{k}^{a},\;E_{l}^{b}\right\} =\delta ^{ab}\delta _{k}^{l}.
\end{equation}
\subsection{The $E-E$ sector}
We have%
\begin{equation}
\left\{ E_{3}^{1},E_{\bar{\imath}}^{1}\right\} =0,\;\;\;\left\{
E_{3}^{1},E_{i}^{\bar{a}}\right\} =0,\;\;\;\;\;\left\{ E_{\bar{\imath}%
}^{1},E_{\bar{k}}^{1}\right\} =0,\;\;\;\;\;\left\{ E_{\bar{\imath}%
}^{1},E_{k}^{\bar{a}}\right\} =0,\;\;\;\left\{ E_{k}^{\bar{a}},E_{l}^{\bar{b}%
}\right\} =0.
\end{equation}%
To summarize, the PB algebra of the canonical variables in the YM theory,
calculated from the canonical algebra of the corresponding SL-NANM,
reproduces the YM algebra (\ref{FINALDB}). 

\subsection{The $\left( A_{0}^{a},\;\Pi _{0}^{a}\right) -\left(
A_{i}^{a},\;E_{k}^{b}\right) $ sector}

We also need the PBs between the $A_{0}^{a},\;\Pi _{0}^{a}$ and the YM variables 
$A_{i}^{a},\;E_{k}^{b}$. The results are%
\begin{equation}
\left\{ A_{0}^{a},\;A_{i}^{b}\;\right\} =0,\;\;\;\;\;\;\left\{
A_{0}^{a},\;E_{j}^{b}\right\} =\frac{A_{j}^{b}}{A_{0}^{1}}\delta
^{1a},\;\;\;\;  \label{PBA0E1}
\end{equation}%
\begin{eqnarray}
\left\{ \Pi _{0}^{a},\;A_{i}^{b}\;\right\} &=&-\delta ^{1b}\delta _{i}^{3}%
\frac{A_{0}^{a}}{A_{3}^{1}},\qquad \left\{ \Pi _{0}^{a},\;E_{3}^{1}\right\} =%
\frac{E_{3}^{1}}{A_{0}^{1}}\left[ \delta ^{a1}-\frac{A_{0}^{a}A_{0}^{1}}{%
\left( A_{3}^{1}\right) ^{2}}\right] ,  \nonumber \\
\left\{ \Pi _{0}^{a},\;E_{\bar{\imath}}^{1}\right\} &=&\frac{E_{3}^{1}}{%
A_{3}^{1}}\frac{A_{\bar{\imath}}^{1}}{A_{0}^{1}}\delta ^{a1},\qquad \left\{
\Pi _{0}^{a},\;E_{i}^{\bar{b}}\right\} =\frac{E_{3}^{1}}{A_{3}^{1}}\frac{%
A_{i}^{\bar{b}}}{A_{0}^{1}}\delta ^{a1}.  \label{PBPI0R}
\end{eqnarray}

\subsection{The $\Phi _{1}^{\bar{a}}-(A_{i}^{a},\;A_{0}^{a},\;E_{i}^{a})$
sector}
The equality of the Dirac brackets and the PBs derived in Eq. (\ref{DBEPBSL}) is
a direct consequence of the PBs between the constraints $\Phi _{1}^{\bar{a}%
}=\Pi _{0}^{\bar{a}}-\Pi _{0}^{1}A_{0}^{\bar{a}}/A_{0}^{1}$ \ and the YM
variables, which are 
\begin{eqnarray}
\left\{ \Phi _{1}^{\bar{a}},E_{3}^{1}\;\right\} &=&0,\;\;\;\;\left\{ \Phi
_{1}^{\bar{a}},E_{\bar{\imath}}^{1}\;\right\} =0,\;\;\;\left\{ \Phi _{1}^{%
\bar{a}},E_{k}^{\bar{b}}\;\right\} =0,\;\left\{ \Phi _{1}^{\bar{a}%
},E_{k}^{b}\;\right\} =0,  \nonumber \\
\;\;\left\{ \Phi _{1}^{\bar{a}},\;A_{0}^{1}\;\right\} &=&\frac{A_{0}^{\bar{a}%
}}{A_{0}^{1}},\;\;\;\;\left\{ \Phi _{1}^{\bar{a}},\;A_{0}^{\bar{b}%
}\;\right\} =-\delta ^{\bar{a}\bar{b}},\;\;\left\{ \Phi _{1}^{\bar{a}%
},\;\;A_{3}^{1}\right\} =0,\;\;\;\left\{ \Phi _{1}^{\bar{a}},\;\;A_{\bar{%
\imath}}^{1}\right\} =0,\;\;\;\;  \nonumber \\
\left\{ \Phi _{1}^{\bar{a}},\;\;A_{k}^{\bar{b}}\right\} &=&0,\;\;\;\left\{
\Phi _{1}^{\bar{a}},\;\;A_{k}^{a}\right\} =0,\;\;\;\left\{ \Phi _{1}^{\bar{a}%
},\;B_{k}^{b}\;\right\} =0.  \label{CPHI11}
\end{eqnarray}

\subsection{The constraints sector}

Next we provide the results for the calculation of the PBs of the
constraints $\Phi _{1}^{\bar{a}}$ and $\Phi _{2}^{\bar{b}}$: 
\begin{equation}
\left\{ \Phi _{1}^{\bar{a}},\;\;\Phi _{1}^{\bar{b}}\right\} =0.
\label{PBPHI1PHI1}
\end{equation}%
The calculation of $\left\{ \Phi _{1}^{\bar{a}},\;\Phi _{2}^{\bar{b}%
}\right\} $ requires 
\begin{equation}
\left\{ \Phi _{1}^{\bar{a}},\;\Omega ^{b}\right\} =0,  \label{PH1OMEGA}
\end{equation}%
in virtue of the relation $\left\{ \Phi _{1}^{\bar{a}},\;E_{k}^{b}\right\}
=0 $ calculated in (\ref{CPHI11}). The final result is%
\begin{equation}
\left\{ \Phi _{1}^{\bar{a}},\;\Phi _{2}^{\bar{b}}\right\} =-\left( \delta ^{%
\bar{a}\bar{b}}+\frac{A_{0}^{\bar{a}}}{A_{0}^{1}}\frac{A_{0}^{\bar{b}}}{%
A_{0}^{1}}\right),  \label{PBPHI1PHI2}
\end{equation}%
where we also have made use of the constraint $\Omega ^{1}A_{0}^{\bar{a}%
}=A_{0}^{1}\Omega ^{\bar{a}}$ from Eq. (\ref{SCC_NANM}).

Using the results 
\begin{equation}
\left\{ A_{0}^{1},\;\;\Omega ^{b} \right\} =0,\;\;\;\;\;\;\left\{ A_{0}^{%
\bar{a}},\;\;\Omega ^{b} \right\} =0,
\end{equation}%
we calculate 
\begin{equation}
\left\{ \Phi _{2}^{\bar{a}},\;\Phi _{2}^{\bar{b}}\right\} =\left(
A_{0}^{1}\right) ^{2}\left\{ \frac{\Omega ^{\bar{a}}}{\Omega ^{1}},\frac{%
\Omega ^{\bar{b}}}{\Omega ^{1}}\;\right\} .
\end{equation}%
The last PB is 
\begin{equation}
\left\{ \frac{\Omega ^{\bar{a}}}{\Omega ^{1}},\frac{\Omega ^{\bar{b}}}{%
\Omega ^{1}}\;\right\} =\frac{1}{\left( \Omega ^{1}\right) ^{2}}\left( C^{%
\bar{a}\bar{b}m}+C^{1\bar{a}m}\frac{\Omega ^{\bar{b}}}{\Omega ^{1}}-C^{1\bar{%
b}m}\frac{\Omega ^{\bar{a}}}{\Omega ^{1}}\right) D_{i}E_{i}^{m},
\end{equation}%
where the relation%
\begin{equation}
\;\;\left\{ \Omega ^{a},\;\;\Omega ^{b}\right\} =C^{abc}D_{i}E_{i}^{c},
\label{PPOMEGA}
\end{equation}%
has been used. \ The final result is 
\begin{equation}
\left\{ \Phi _{2}^{\bar{a}},\;\Phi _{2}^{\bar{b}}\right\} =\left( \frac{%
A_{0}^{1}}{\Omega ^{1}}\right) ^{2}\left( C^{\bar{a}\bar{b}m}+C^{1\bar{a}m}%
\frac{\Omega ^{\bar{b}}}{\Omega ^{1}}-C^{1\bar{b}m}\frac{\Omega ^{\bar{a}}}{%
\Omega ^{1}}\right) \left( D_{i}E_{i}^{m}\right) .  \label{PPPHI2PHI2}
\end{equation}

\section{The Wronskian}

In the formulation of the NANM presented in Sec. III,  the Wronskian
arising from the Lagrangian is 
\begin{equation}
W_{\rm NANM}=\det \left( \frac{\partial A_{i}^{c}}{\partial \Phi _{B}^{b}}\frac{%
\partial A_{i}^{c}}{\partial \Phi _{A}^{a}}\right)
,\;\;\;i,A=1,2,3,\;\;\;\;\;\;\;a=1,...,N,  \label{W}
\end{equation}%
where the coordinate transformation $A_{i}^{a}=A_{i}^{a}(\Phi _{B}^{b})$ is  invertible so that 
\begin{equation}
\;\det \left( \frac{\partial A_{i}^{a}}{\partial \Phi _{B}^{b}}\right) \neq
0.  \label{INVERTA3}
\end{equation}%
We next show that the property (\ref{INVERTA3}) guarantees that $W_{\rm NANM}$ is nonzero. To this end, it is simpler to relabel the coordinate
transformation from $3N$ $A_{i}^{a}\;$to $\;3N$ $\Phi _{B}^{b}\;$as%
\begin{equation}
A_{R}=A_{R}(\Phi _{S}),\;\;\;\det \left( \frac{\partial A_{R}}{\partial \Phi
_{S}}\right) \neq 0\;,\;\;\;R,S=1,2,3,...3N\;  \label{INVERTGEN}
\end{equation}%
The required Wronskian is 
\begin{equation}
W_{\rm NANM}=\det \left( \frac{\partial A_{R}}{\partial \Phi _{S}}\frac{\partial A_{R}}{%
\partial \Phi _{T}}\right) .
\end{equation}%
In terms of the $3N$-dimensional epsilon symbol $\epsilon
^{T_{1}T_{2},...,T_{3N}}$, the above can be written as 
\begin{eqnarray}
W_{\rm NANM} &=&\epsilon ^{T_{1}T_{2},...,T_{3N}}\left( \frac{\partial A_{R_{1}}}{%
\partial \Phi _{1}}\frac{\partial A_{R_{1}}}{\partial \Phi _{T_{1}}}\right)
\left( \frac{\partial A_{R_{2}}}{\partial \Phi _{2}}\frac{\partial A_{R_{2}}%
}{\partial \Phi _{T_{2}}}\right) ...\left( \frac{\partial A_{R_{3N}}}{%
\partial \Phi _{3N}}\frac{\partial A_{R_{3N}}}{\partial \Phi _{T_{3N}}}%
\right) .  \nonumber \\
W_{\rm NANM} &=&\left( \frac{\partial A_{R_{1}}}{\partial \Phi _{1}}\frac{\partial
A_{R_{2}}}{\partial \Phi _{2}}...\frac{\partial A_{R_{3N}}}{\partial \Phi
_{3N}}\right) \left( \epsilon ^{T_{1}T_{2},...,T_{3N}}\frac{\partial
A_{R_{1}}}{\partial \Phi _{T_{1}}}\frac{\partial A_{R_{2}}}{\partial \Phi
_{T_{2}}}....\frac{\partial A_{R_{3N}}}{\partial \Phi _{T_{3N}}}\right) .
\end{eqnarray}%
But%
\begin{equation}
\left( \epsilon ^{T_{1}T_{2},...,T_{3N}}\frac{\partial A_{R_{1}}}{\partial
\Phi _{T_{1}}}\frac{\partial A_{R_{2}}}{\partial \Phi _{T_{2}}}....\frac{%
\partial A_{R_{3N}}}{\partial \Phi _{T_{3N}}}\right) =\epsilon
^{R_{1}R_{2},...,R_{3N}}\det \left( \frac{\partial A_{R}}{\partial \Phi _{T}}%
\right) ,
\end{equation}%
so that 
\begin{eqnarray}
W_{\rm NANM} &=&\left( \frac{\partial A_{R_{1}}}{\partial \Phi _{1}}\frac{\partial
A_{R_{2}}}{\partial \Phi _{2}}...\frac{\partial A_{R_{3N}}}{\partial \Phi
_{3N}}\right) \epsilon ^{R_{1}R_{2},...,R_{3N}}\det \left( \frac{\partial
A_{R}}{\partial \Phi _{T}}\right) ,  \nonumber \\
W_{\rm NANM} &=&\left[ \det \left( \frac{\partial A_{R}}{\partial \Phi _{T}}\right) %
\right] ^{2}\neq 0,
\end{eqnarray}%
employing (\ref{INVERTGEN}).

\end{document}